\documentclass[iop]{emulateapj}

\usepackage{natbib}

\newcommand{\kms}{km\,s$^{-1}$}

\newcommand{\msun}{$M_\sun$}

\newcommand{\sqdeg}{deg$^2$}

\newcommand{\mum}{$\mu$m}
\newcommand{\cmpc}{Mpc$^{-3}$}
\newcommand{\mbh}{$M_\mathrm{BH}$}
\newcommand{\ks}{$K_\mathrm{s}$}

\slugcomment{Received 2013 July 11; accepted 2013 September 27; published 2013 November 22}

\shorttitle{Discovery of three $z>6.5$ quasars in VIKING} 
\shortauthors{Venemans et al.}

\begin{document}

\title{DISCOVERY OF THREE $z>6.5$ QUASARS IN THE VISTA KILO-DEGREE INFRARED
  GALAXY (VIKING) SURVEY\altaffilmark{1}}

\author{B. P. Venemans\altaffilmark{2}, J. R. Findlay\altaffilmark{3}, W. J. Sutherland\altaffilmark{4}, G. De Rosa\altaffilmark{5,6}, R. G. McMahon\altaffilmark{7,8}, R. Simcoe\altaffilmark{9}, E. A. Gonz\'alez-Solares\altaffilmark{7}, K. Kuijken\altaffilmark{10}, and J. R. Lewis\altaffilmark{7}}

\altaffiltext{1}{Based on observations collected at the European
    Southern Observatory, Chile, programs 179.A-2004, 087.A-0717, 087.A-0890
    and 088.A-0897. This paper also includes data gathered with the 6.5 meter
    Magellan Telescopes located at Las Campanas Observatory, Chile.}
\altaffiltext{2}{Max-Planck Institute for Astronomy, K{\"o}nigstuhl 17, 69117
  Heidelberg, Germany; venemans@mpia.de}
\altaffiltext{3}{Department of Physics, Durham University, South
  Road, Durham, DH1 3LE, UK}
\altaffiltext{4}{Astronomy Unit, School of Mathematical Sciences, Queen Mary, University
of London, London, E1 4NS, UK}
\altaffiltext{5}{Department of Astronomy, The Ohio State University,
  140 West 18th Avenue, Columbus, OH 43210, USA}
\altaffiltext{6}{Center for Cosmology \& AstroParticle Physics, The Ohio State University, 191 West Woodruff Ave, Columbus, OH 43210, USA}
\altaffiltext{7}{Institute of Astronomy, University of Cambridge,
  Madingley Road, Cambridge, CB3 0HA, UK}
\altaffiltext{8}{Kavli Institute for Cosmology, University of Cambridge, Madingley Road,
  Cambridge CB3 0HA, UK}
\altaffiltext{9}{MIT-Kavli Center for Astrophysics and Space
  Research, 77 Massachusetts Avenue, Cambridge, MA 02139, USA}
\altaffiltext{10}{Leiden Observatory, Leiden University, Niels
  Bohrweg 2, NL-2333 CA Leiden, The Netherlands}

\begin{abstract}
  Studying quasars at the highest redshifts can constrain models of galaxy and
  black hole formation, and it also probes the intergalactic medium in the early
  universe. Optical surveys have to date discovered more than 60 quasars up to
  $z\simeq6.4$, a limit set by the use of the $z$-band and CCD detectors. Only
  one $z\gtrsim6.4$ quasar has been discovered, namely the $z=7.08$ quasar
  ULAS J1120+0641, using near-infrared imaging. Here we report the discovery
  of three new $z\gtrsim6.4$ quasars in 332\,\sqdeg\ of the Visible and Infrared Survey Telescope for Astronomy Kilo-degree
  Infrared Galaxy (VIKING) survey, thus extending the number from 1 to 4. The
  newly discovered quasars have redshifts of $z=6.60$, $6.75$, and $6.89$. The
  absolute magnitudes are between --26.0 and --25.5, 0.6--1.1\,mag fainter
  than ULAS J1120+0641. Near-infrared spectroscopy revealed the \ion{Mg}{2}
  emission line in all three objects. The quasars are powered by black holes
  with masses of $\sim$$(1-2)\times10^9$\,\msun. In our probed redshift range
  of $6.44<z<7.44$ we can set a lower limit on the space density of
  supermassive black holes of $\rho(M_\mathrm{BH}>10^9\,M_\odot) >
  1.1\times10^{-9}$\,\cmpc. The discovery of three quasars in our survey area
  is consistent with the $z=6$ quasar luminosity function when extrapolated to
  $z\sim7$. We do not find evidence for a steeper decline in the space density
  of quasars with increasing redshift from $z=6$ to $z=7$.
\end{abstract}


\section{INTRODUCTION}

The cosmic microwave background reveals the earliest observable
structure in the universe, which dates back to the epoch of
recombination at $z \sim 1100$. The next most distant observable structure
represents the population of galaxies and quasars at $z\sim7-10$. The
interval between $z=1000$ and $z=7$ contains many landmark events,
such as the formation of the first stars and galaxies, the growth of
the first massive black holes and the reionization of the neutral
hydrogen in the intergalactic medium (IGM). Measurements by the {\it
  Wilkinson Microwave Anisotropy Probe} give a $2\,\sigma$ lower bound
to the epoch of reionization of $z_r=8.2$ \citep{kom11} but do not
pin down when reionization began or ended. Another probe of the
ionized state of the IGM are spectra of high redshift
quasars. Already, the analysis of spectra of $z\sim6$ quasars suggest
that the epoch of reionization ended around $z\sim6.1$
\citep{gne06,bec07}.

While in recent years the evidence is pointing toward galaxies providing the
required UV flux to reionize the universe \citep[e.g.][]{wil10a,bou12}, the
identification and study of quasars up to the highest redshifts is very
important for several reasons: (1) absorption line studies of cosmologically
distributed intervening material allow the determination of the baryonic
content and physical conditions (metallicity, temperature, ionization state)
of the universe during the epoch of reionization
\citep[e.g.][]{mor11,bol11,sim12}; (2) the number density of high redshift
quasars provide constraints on the mechanisms that are required to seed and
grow $>10^9$\,\msun\ supermassive black holes less than a Gyr after the big
bang \citep[e.g.,][]{lat13}; and (3) high redshift quasars are generally
located in luminous, massive host galaxies. By studying the host galaxies we
get insight into the formation of galaxies in the early universe
\citep[e.g.,][]{wal09b,wan13}.

In the last decade more than sixty $5.7<z\lesssim6.4$ quasars have been
discovered in various different surveys
\citep[e.g.,][]{fan06b,ven07b,jia08,jia09,wil10a,mor12}. The vast majority of
these quasars have been discovered in optical surveys, in particular the Sloan
Digital Sky Survey \citep[SDSS; e.g.,][]{fan06b} and the Canada--France High--$z$ Quasar
Survey \citep[CFHQS;][]{wil10a}. The limitation of optical surveys that use
silicon based CCD detectors is that they cannot find quasars beyond $z\sim6.4$
as the sources become too faint in the reddest optical band of the survey
($z$-band) due to absorption by the intervening Ly$\alpha$ forest. To overcome
this limit large area surveys in near-infrared bandpasses are necessary.

One of the first wide-field near-infrared surveys reaching a depth that
allowed a search for the highest redshift quasars was the UK Infrared
Telescope Infrared Deep Sky Survey (UKIDSS) Large Area Survey
\citep[LAS;][]{law07}. Using the UKIDSS LAS, \citet{mor11} discovered the
first quasar beyond $z>6.5$, ULAS J1120+0641 at $z=7.1$. This one
$z\sim7$ quasar already provides constraints on the neutral fraction of hydrogen up to
$z=7.1$ \citep{bol11,mor11}, on the metallicity of the IGM \citep{sim12}, and
on the formation of dust and stars in the host galaxy \citep{ven12}. This
shows that more quasars at $z>6.5$ are needed. 

Here we present the first $z>6.5$ quasars discovered in the Visible and
Infrared Survey Telescope for Astronomy (VISTA) Kilo-Degree
Infrared Galaxy (VIKING) survey. In Section \ref{surveydata} we give a short
description of the VIKING survey. In Section \ref{candidateselection} we
discuss the selection of high-redshift quasar candidates in the VIKING data,
followed by the details of the imaging and spectroscopic observations of
quasar candidates in Section \ref{followup}. We continue with an overview of
the results of the follow-up observations in Section \ref{results}. In
Section \ref{properties} we characterize the newly discovered quasars. In
Sections \ref{catcomp}--\ref{colcomp} we assess various sources of
incompleteness and the effect on our quasar search. We discuss the
implications of our results on the quasar luminosity function at $z\sim7$ in
Section \ref{qsolf}. We conclude with a summary in Section
\ref{discussion}.

All magnitudes are given in the AB system. For magnitudes obtained
with VISTA we use the Vega to AB conversions provided on the Cambridge
Astronomical Survey Unit (CASU)
website\footnote{\url{http://casu.ast.cam.ac.uk/}}:
$Z_{\mathrm{AB}}=Z_{\mathrm{Vega}}+0.521$,
$Y_{\mathrm{AB}}=Y_{\mathrm{Vega}}+0.618$,
$J_{\mathrm{AB}}=J_{\mathrm{Vega}}+0.937$,
$H_{\mathrm{AB}}=H_{\mathrm{Vega}}+1.384$ and
$K_{\mathrm{s,AB}}=K_{\mathrm{s,Vega}}+1.839$.  An $\Lambda$-dominated
cosmology with $\Omega_M=0.28$, $\Omega_{\Lambda}=0.72$ and
$H_0=70$\,km\,s$^{-1}$\,Mpc$^{-1}$ \citep{kom11} is adopted.

\section{SURVEY DATA}
\label{surveydata}

The VIKING survey is one of six public surveys \citep{arn07} ongoing since
late 2009 on VISTA, a 4\,m wide field survey telescope located at ESO's Cerro
Paranal Observatory in Chile \citep{eme06}. The telescope is equipped with the
near-infrared camera VISTA InfraRed CAMera \citep{dal06}, which
contains sixteen 2048$\times$2048\,pixel$^2$ Raytheon VIRGO HgCdTe infrared
detectors and has a field of view of 1.65\,deg in diameter. With a pixel scale
of 0\farcs34\,pixel$^{-1}$, the instantaneous field of view (called a
``pawprint'') is 0.6\,\sqdeg. Because the detectors sparsely sample the field
of view, a mosaic of six ``pawprints'' carried out in a six-step pattern covers
a continuous area of 1.5\,\sqdeg\ of the sky (a ``tile'').

The VIKING survey aims to cover 1500\,\sqdeg\ spread over three
extragalactic areas in five broad-band filters --- $Z$, $Y$, $J$, $H$,
and \ks. The details of the five broad-band filters and the limiting
magnitudes of the VIKING survey are listed in Table
\ref{vikingfilters}. The three areas and the R.A.\ and decl.\ limits
are as follows:

\begin{figure}[t]
\epsscale{1.2}
\plotone{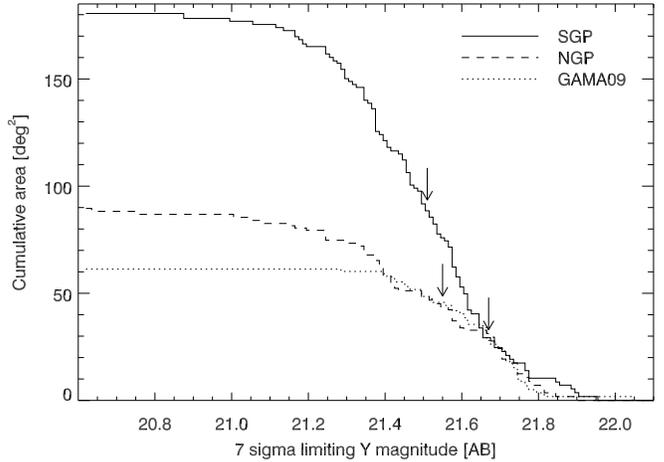}
\caption{\footnotesize Cumulative area in square degrees covered by the VIKING survey in the
  $Y$-band up to 2011 March as a function of $7\,\sigma$ limiting magnitude. The
  median depth for each region is indicated by an arrow. \label{areadepth}}
\end{figure}

\begin{deluxetable}{lcccc}
\tablecaption{Characteristics of the Filters Used in the VIKING Survey and
  the Nominal 5$\sigma$ Limiting Magnitudes \label{vikingfilters}} 
\tablewidth{0pt} 
\tablehead{ \colhead{Filter} & \colhead{Central Wavelength\tablenotemark{a}} &
  \colhead{Width\tablenotemark{a}} & \colhead{Exp.\ Time\tablenotemark{b}} & 
  \colhead{Depth\tablenotemark{c}}}
\startdata
$Z$ & 0.877 & 0.097 & 480 & 23.1 \\
$Y$ & 1.020 & 0.093 & 400 & 22.3 \\
$J$ & 1.252 & 0.172 & 400 & 22.1 \\
$H$ & 1.645 & 0.291 & 300 & 21.5 \\
\ks & 2.147 & 0.309 & 480 & 21.2 \\
\enddata
\tablenotetext{a}{Filter central wavelength and width in \mum.}
\tablenotetext{b}{Typical exposure time per pixel in s.}
\tablenotetext{c}{5$\sigma$ AB magnitudes.}
\end{deluxetable}

\noindent
1. SGP: 22$^h$00$^m$\,$<$\,R.A.\,$<$\,03$^h$30$^m$,
$-36^\circ$\,$<$\,decl.\,$<$\,$-26^\circ$; \\
\noindent
2. NGP: 10$^h$30$^m$\,$<$\,R.A.\,$<$\,15$^h$30$^m$,
$-5^\circ$\,$<$\,decl.\,$<$\,$+4^\circ$; \\
\noindent
3. GAMA09: 08$^h$36$^m$$<$R.A.$<$09$^h$24$^m$,
$-2^\circ$$<$decl.$<$$+3^\circ$.

The data are processed by CASU and archived at the Wide Field Astronomy Unit
at the Royal Observatory in Edinburgh as part of the VISTA Data Flow
System \citep{lew05,lew10}. At CASU all the VISTA data are processed
to science products, including astrometric and photometric calibration, for
each night of data \citep{irw04}. The stacking of pawprints into a tile is
done if a suitable set of pawprints has been taken within a single Observing
Block (OB). For the VIKING survey, the observations of a single tile in the
five broad-band filters are spread over two OBs: in one OB, the $Z$-, $Y$- and
half the $J$-band are taken, while the second OB consists of the $H$-, and
\ks-band images and the second half of the $J$-band observations. After the
reduction at CASU, we thus have $Z$-, $Y$-, $H$- and \ks-band images and
catalogs to the full depth, while we have two separate $J$-band images with
half the total exposure time.

The VIKING data used in this article were obtained between 2009 November and
2011 March. We downloaded the processed images and catalogs directly from CASU
and band-merged the catalogs ourselves. When we had two separate $J$-band
catalogs for the same tile, we merged the two catalogs. A search radius of
1\farcs5 was used to merge the catalogs. The search radius is roughly three
times that of the seeing measured in the VIKING data, which has a medium FWHM
of 0\farcs99. If no entry was found in a catalog, a $3\,\sigma$ limit was
assigned to the object. The total number of tiles observed up to 2011 March
for which we had at least $Z$-, $Y$- and $J$-band catalogs was 44 in the
GAMA09 region, 59 in the NGP and 121 in the SGP. The total unique area
covered by these tiles was 331.6\,\sqdeg\ (61.2\,\sqdeg\ in GAMA09,
89.6\,\sqdeg\ in the NGP and 180.8\,\sqdeg\ in the SGP). The area as function
of $7\,\sigma$ depth in the $Y$-band is shown in Figure \ref{areadepth}. The
median $7\,\sigma$ $Y$-band depths are 21.67, 21.55, and 21.51 for the GAMA09,
NGP, and SGP regions respectively.

\section{CANDIDATE SELECTION}
\label{candidateselection}

We selected candidate high redshift ($z\gtrsim6.5$) quasars from the
band-merged catalogs\footnote{For a detailed description of all the parameters
  in the catalogs that are generated at CASU, see
  \url{http://casu.ast.cam.ac.uk/surveys-projects/vista/technical/catalogue-generation}}
by applying the following steps:

\begin{enumerate}

\item Objects should be detected in the $Y$-band with a signal-to-noise ratio (S/N)
  of at least (S/N)$_Y > 7$. We also required that the object is detected in at
  least one other band. For all bands we use aperture-corrected magnitudes
  measured in an aperture with a radius of 1\farcs0 (labeled aperMag3 in the
  catalogs).

\item We only consider point sources that we define as objects having
  a probability of being a galaxy (pGalaxy) of pGalaxy\,$< 0.95$
  \citep[see][ for details on the morphological
  parameters]{irw04,gon08}. This selection makes sure that most real
  point sources are selected with a manageable fraction of galaxies
  \citep[see also Section \ref{pscomp} and][]{fin12}.

\item The aperture-corrected magnitudes in the $Y$-band had to be independent
  of the size of aperture used to within 0.2\,mag. This was forced by applying
  aperMag1$-$aperMag3\,$< 0.2$ and aperMag3$-$aperMag4\,$< 0.2$. Removing this
  criterion did not increase the number of candidate quasars, but did add
  more artifacts to our candidate lists. 

\item Because $z>6.5$ quasars occupy a unique region in color space, we can
  isolate potential quasars in the catalogs by introducing color criteria. We
  applied two sets of color criteria to select $z>6.5$ quasar candidates, a
  conservative color selection and an extended color selection. The
  conservative color selection minimizes the number of foreground objects
  wrongly selected as candidate quasars and is based on the work of
  \citet{fin12}. The extended color criteria maximize the completeness of the
  high redshift quasar selection, without purposely selecting foreground
  objects. The conservative color criteria are as follows:
  
  $Z-Y \geq 1.25$ AND $-0.5 < Y-J \leq 0.5$ AND ($Z > Z_{\mathrm{lim},
    3\sigma}$ OR $Z-Y\,>\,0.75\times(Y-J)\,+\,1.45$) AND $-0.5 < Y-K < 1.3$
  AND undetected in $u$, $g$, $r$ and $i$ (when optical imaging is available). 

  \noindent
  The extended color criteria are:

  $Z-Y \geq 1.1$ AND $-0.5 < Y-J \leq 0.5$ AND ($Z > Z_{\mathrm{lim}, 3\sigma}$ OR
  $Z-Y\,>\,Y-J\,+\,0.7$) AND $-0.5 < Y-K < 1.0$ AND $J-K < 0.8$ AND undetected
  in $u$, $g$, $r$, $i$ (when optical imaging is available).
  
  \noindent
  Our color criteria are illustrated in Figure \ref{cc_zyj}.

\item Detector 16 has a time-varying quantum efficiency. As a result the flat
  fielding is not accurate and the detector produces a large number spurious
  detections. We therefore ignore objects that originated from detector 16 in
  one pawprint and were not independently confirmed in another pawprint.

\item To check the magnitudes listed in the catalogs and especially
  nondetections in the $Z$-band, we performed aperture photometry in each of
  the VIKING images at the location of the candidate. If the colors measured
  did not fulfill the color criteria listed above, the object was removed from
  our candidate list. 

\begin{figure}[t]
\epsscale{1.2}
\plotone{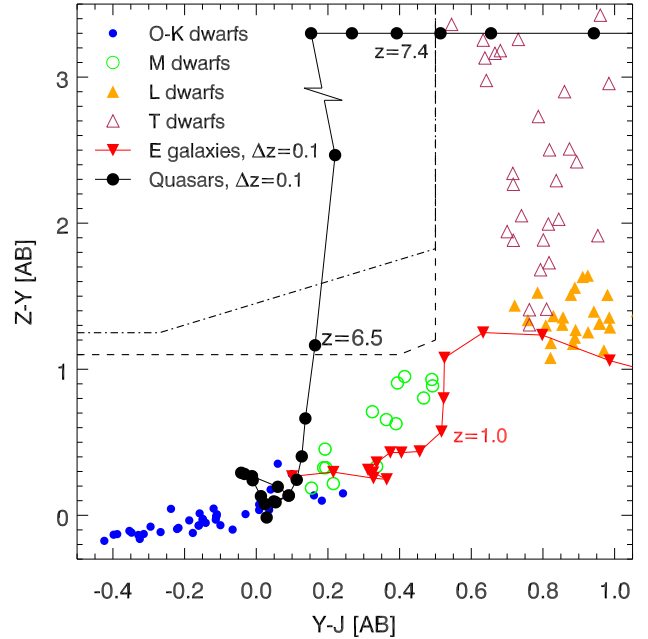}
\caption{$Z-Y$ vs.\ $Y-J$ diagram illustrating our color selection
  criteria. The dashed line shows our extended criteria, while the dash-dotted
  line illustrates our conservative color criteria. Also plotted are simulated
  colors of main sequence stars (small, blue dots), M dwarfs (green, open
  circles), L dwarfs (yellow, solid triangles), T dwarfs (open, maroon
  triangles), elliptical galaxies (red, upside-down triangles), and quasars
  (black, solid points) from \citet{hew06}. Quasars above $z\gtrsim6.4-6.5$
  occupy a region in the color--color diagram that is well separated from
  foreground objects. Quasars above $z>6.6$ become quickly very faint in the
  $Z$-band and have $Z-Y\gg3$. We truncated the quasar track between $z=6.6$
  and $z=7.2$ to show the evolution of the $Y-J$ color at the highest
  redshifts. At redshifts $z\gtrsim7.4-7.5$ quasars become too red in $Y-J$
  color to separate them from brown dwarfs in color space. \label{cc_zyj}}
\end{figure}

\item The final step was visually inspecting all candidates in order
  to remove any remaining spurious objects. Most of the eliminated objects
  were near the edge or just outside the $Z$-band image, or on diffraction
  spikes. This step approximately halved the number of candidates on our lists.

\end{enumerate}

With the conservative color criteria we selected 43 objects in the three
different regions observed by VIKING: 7 objects in GAMA09, 18 in
the NGP, and 18 in the SGP. We applied our extended color selection to the NGP
region and further selected another 42 objects. We loosely ranked the objects
according to their $Z-Y$ color, i.e., objects with a higher $Z-Y$ had a higher
priority for follow-up observations.

\section{FOLLOW-UP OBSERVATIONS}
\label{followup}

\subsection{ESO NTT/EFOSC2 Imaging} 

On 2011 June 26--29 we observed quasar candidates from VIKING with the
European Southern Observatory's Faint Object Spectrograph and Camera 2
\citep[EFOSC2;][]{buz84} on the 3.58 m ESO New Technology Telescope (NTT). We
observed in filters Gunn $i$ (filter \#705, hereafter $I_N$) and Gunn $z$
(filter \#623, hereafter $Z_N$).

Images taken with the $Z_N$ filter were calibrated using stars within the
EFOSC2 field of view. Using the filter transmission curves, the CCD efficiency
and stellar spectra from the BPGS spectroscopic
atlas\footnote{\url{http://www.stsci.edu/hst/observatory/cdbs/bpgs.html}}
\citep{gun83}, we established a relation between the $Z_N$ and the VIKING $Z$
and $Y$ magnitudes of $Z_N=Z-0.135\times(Z-Y)+0.01$. In fields with SDSS
coverage (the NGP and GAMA09 regions), the $I_N$ images were calibrated using
the relation (derived from synthesized colors of the BPGS stars)
$I_N=i_\mathrm{SDSS}-0.4\times(i_\mathrm{SDSS}-z_\mathrm{SDSS})$. Observations
in the $I_N$ filter of candidates in the SGP, for which no optical photometry
of nearby stars was available, were calibrated using the spectrophotometric
standard stars LTT1020, LTT6248, and LTT9239 \citep{ham92,ham94} that were
observed in the evening and morning twilight.

During the NTT run we observed 18 quasar candidates that were selected with
the conservative color criteria and 20 sources that fulfilled the extended color
criteria. To fill gaps in the nights, we also targeted seven objects that fell just
outside the color selection and/or had a $Y$-band detection just below the
(S/N)$_Y$\,$=7$ limit. 

\begin{figure}[t]
\epsscale{1.2}
\plotone{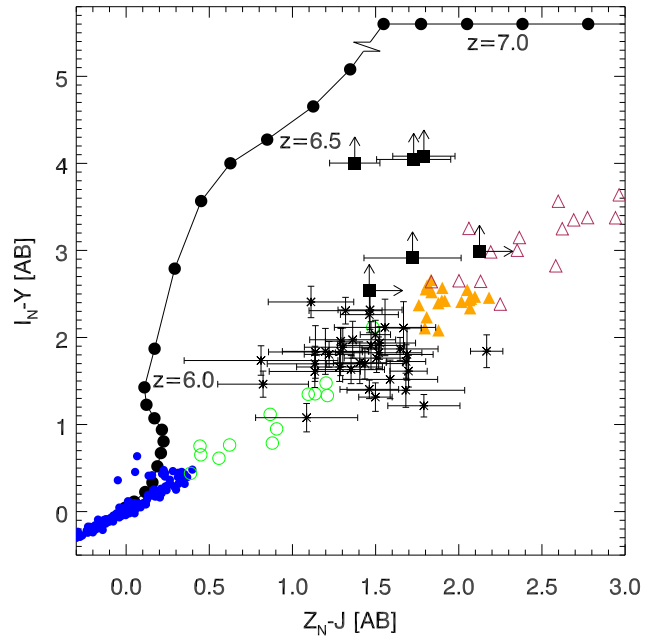}
\caption{$I_N-Y$ vs.\ $Z_N-J$ diagram showing the results of the follow-up
  observations with the NTT. Similar to Figure \ref{cc_zyj}, colors of
  simulated stars, galaxies, and quasars are also shown (see the caption of
  Figure \ref{cc_zyj} for an explanation of the symbols). The value of the
  $I_N-Y$ color of quasars above $z>6.7$ have been reduced to $I_N-Y=5.6$ in
  order to display the change in $Z_N-J$ color of quasars with redshift more
  clearly. The small stars with error bars represent candidates detected in
  the $I_N$ images and were identified as foreground stars. The six filled
  squares are candidates that were not detected in the $I_N$ images and
  therefore remained good high-redshift quasar candidates. These objects were
  our prime targets for follow-up spectroscopy. \label{nttccplot}}
\end{figure}

Individual images had an exposure time of either 600\,s or 900\,s, and the
total exposure time per source varied from 600\,s to 8100\,s. We observed 45
objects in the $I_N$ filter, and 8 of those were also observed in the $Z_N$
filter. Generally, if an object showed a clear detection in the initial $I_N$
images, we stopped observing it. Objects not detected in $I_N$ were
prioritized for deeper $I_N$ and $Z_N$ imaging. The results of the EFOSC2
photometry are listed in Table \ref{efoscobs} and are shown in Figure
\ref{nttccplot}.

\begin{deluxetable*}{lcccc}
\tablecaption{Photometry Obtained with EFOSC2 on the NTT  \label{efoscobs}}
\tablewidth{0pt}
\tablehead{
\colhead{Name} & \colhead{R.A. (J2000)} & \colhead{Decl.(J2000)} &
  \colhead{$I_{N,\mathrm{AB}}$} & \colhead{$Z_{N,\mathrm{AB}}$}
}
\startdata
J0109--3047 &  01$^h$09$^m$53$^s\!\!$.13 & --30$^\circ$47\arcmin26\farcs3 & $>$25.27 & 22.93$\pm$0.16 \\
J0125--3016 &  01$^h$25$^m$34$^s\!\!$.44 & --30$^\circ$16\arcmin27\farcs4 & 22.82$\pm$0.11 & -- \\
J0221--3139 &  02$^h$21$^m$46$^s\!\!$.08 & --31$^\circ$39\arcmin49\farcs4 & 22.35$\pm$0.08 & -- \\
J0223--3208 &  02$^h$23$^m$31$^s\!\!$.00 & --32$^\circ$08\arcmin37\farcs7 & 22.73$\pm$0.10 & -- \\
J0223--3144 &  02$^h$23$^m$53$^s\!\!$.26 & --31$^\circ$44\arcmin26\farcs6 & 23.37$\pm$0.26 & -- \\
J0259--3100 &  02$^h$59$^m$42$^s\!\!$.43 & --31$^\circ$00\arcmin54\farcs4 & 22.57$\pm$0.12 & -- \\
J0305--3150 &  03$^h$05$^m$16$^s\!\!$.92 & --31$^\circ$50\arcmin56\farcs0 & $>$24.97 & 22.12$\pm$0.07 \\
J0305--3400 &  03$^h$05$^m$58$^s\!\!$.11 & --34$^\circ$00\arcmin01\farcs9 & 22.56$\pm$0.10 & -- \\
J0322--3350 &  03$^h$22$^m$13$^s\!\!$.93 & --33$^\circ$50\arcmin25\farcs2 & 23.26$\pm$0.24 & -- \\
J0914+0155 &  09$^h$14$^m$19$^s\!\!$.52 &  +01$^\circ$55\arcmin52\farcs0 & 23.70$\pm$0.29 & -- \\
J1146+0145 &  11$^h$46$^m$26$^s\!\!$.56 &  +01$^\circ$45\arcmin48\farcs8 & 22.73$\pm$0.10 & -- \\
J1147+0126 &  11$^h$47$^m$24$^s\!\!$.90 &  +01$^\circ$26\arcmin34\farcs6 & 22.94$\pm$0.12 & -- \\
J1149--0046 &  11$^h$49$^m$35$^s\!\!$.20 & --00$^\circ$46\arcmin00\farcs8 & 22.87$\pm$0.12 & -- \\
J1151+0118 &  11$^h$51$^m$31$^s\!\!$.68 &  +01$^\circ$18\arcmin05\farcs3 & 23.67$\pm$0.22 & -- \\
J1154+0141 &  11$^h$54$^m$39$^s\!\!$.84 &  +01$^\circ$41\arcmin53\farcs3 & $>$24.18 & $>$22.81 \\
J1156--0007 &  11$^h$56$^m$42$^s\!\!$.26 & --00$^\circ$07\arcmin58\farcs1 & $>$24.49 & $>$23.22 \\
J1211+0007 &  12$^h$11$^m$43$^s\!\!$.09 &  +00$^\circ$07\arcmin31\farcs2 & 22.95$\pm$0.14 & -- \\
J1213+0159 &  12$^h$13$^m$21$^s\!\!$.91 &  +01$^\circ$59\arcmin45\farcs7 & 23.34$\pm$0.16 & -- \\
J1213+0200 &  12$^h$13$^m$34$^s\!\!$.80 &  +02$^\circ$00\arcmin32\farcs7 & 23.47$\pm$0.19 & -- \\
J1218+0051 &  12$^h$18$^m$24$^s\!\!$.54 &  +00$^\circ$51\arcmin12\farcs9 & 22.96$\pm$0.15 & -- \\
J1218+0106 &  12$^h$18$^m$46$^s\!\!$.88 &  +01$^\circ$06\arcmin24\farcs2 & 22.65$\pm$0.12 & -- \\
J1219+0003 &  12$^h$19$^m$12$^s\!\!$.11 &  +00$^\circ$03\arcmin26\farcs4 & 22.46$\pm$0.07 & -- \\
J1412+0128 &  14$^h$12$^m$33$^s\!\!$.63 &  +01$^\circ$28\arcmin10\farcs0 & 23.10$\pm$0.09 & -- \\
J1413+0139 &  14$^h$13$^m$41$^s\!\!$.55 &  +01$^\circ$39\arcmin55\farcs5 & 23.41$\pm$0.26 & -- \\
J1414+0126 &  14$^h$14$^m$42$^s\!\!$.14 &  +01$^\circ$26\arcmin19\farcs7 & 23.15$\pm$0.11 & -- \\
J1418--0018 &  14$^h$18$^m$11$^s\!\!$.45 & --00$^\circ$18\arcmin53\farcs1 & 23.07$\pm$0.15 & -- \\
J1418+0109 &  14$^h$18$^m$31$^s\!\!$.42 &  +01$^\circ$09\arcmin33\farcs9 & 23.34$\pm$0.13 & -- \\
J1419--0100 &  14$^h$19$^m$56$^s\!\!$.00 & --01$^\circ$00\arcmin15\farcs3 & 23.38$\pm$0.14 & -- \\
J1425--0054 &  14$^h$25$^m$36$^s\!\!$.77 & --00$^\circ$54\arcmin29\farcs5 & 22.39$\pm$0.06 & -- \\
J1429+0142 &  14$^h$29$^m$27$^s\!\!$.27 &  +01$^\circ$42\arcmin56\farcs9 & 22.94$\pm$0.08 & -- \\
J1431+0029 &  14$^h$31$^m$33$^s\!\!$.72 &  +00$^\circ$29\arcmin00\farcs0 & 23.83$\pm$0.15 & 22.58$\pm$0.14 \\
J1434--0003 &  14$^h$34$^m$09$^s\!\!$.86 & --00$^\circ$03\arcmin45\farcs4 & 23.00$\pm$0.09 & -- \\
J1447+0113 &  14$^h$47$^m$05$^s\!\!$.74 &  +01$^\circ$13\arcmin46\farcs4 & 23.15$\pm$0.10 & -- \\
J1450--0023 &  14$^h$50$^m$12$^s\!\!$.55 & --00$^\circ$23\arcmin45\farcs7 & 23.24$\pm$0.10 & -- \\
J1450+0100 &  14$^h$50$^m$20$^s\!\!$.11 &  +01$^\circ$00\arcmin22\farcs1 & 22.37$\pm$0.05 & -- \\
J1451+0021 &  14$^h$51$^m$37$^s\!\!$.09 &  +00$^\circ$21\arcmin33\farcs4 & 23.23$\pm$0.09 & -- \\
J2214--3134 &  22$^h$14$^m$59$^s\!\!$.28 & --31$^\circ$34\arcmin39\farcs0 & 23.77$\pm$0.13 & -- \\
J2215--3145 &  22$^h$15$^m$36$^s\!\!$.00 & --31$^\circ$45\arcmin44\farcs6 & 20.97$\pm$0.03 & -- \\
J2218--3154 &  22$^h$18$^m$57$^s\!\!$.36 & --31$^\circ$54\arcmin30\farcs0 & $>$24.77 & -- \\
J2222--3129 &  22$^h$22$^m$12$^s\!\!$.24 & --31$^\circ$29\arcmin42\farcs4 & 23.52$\pm$0.19 & -- \\
J2223--3158 &  22$^h$23$^m$28$^s\!\!$.02 & --31$^\circ$58\arcmin35\farcs6 & 23.29$\pm$0.11 & 22.48$\pm$0.14 \\
J2226--3124 &  22$^h$26$^m$58$^s\!\!$.29 & --31$^\circ$24\arcmin45\farcs3 & 23.59$\pm$0.10 & -- \\
J2229--3057 &  22$^h$29$^m$16$^s\!\!$.29 & --30$^\circ$57\arcmin22\farcs8 & 23.64$\pm$0.16 & 23.07$\pm$0.30 \\
J2348--3054 &  23$^h$48$^m$33$^s\!\!$.34 & --30$^\circ$54\arcmin10\farcs0 & $>$25.13 & 22.97$\pm$0.13 \\
J2358--3355 &  23$^h$58$^m$23$^s\!\!$.36 & --33$^\circ$55\arcmin21\farcs1 & 23.03$\pm$0.14 & -- \\
\enddata
\tablecomments{Objects that were not detected in the images have the 3$\sigma$ limiting magnitude. }
\end{deluxetable*}

Of the 45 objects we observed, 6 remained undetected in the $I_N$-band (see
Section \ref{results}). These sources were targets for our follow-up
spectroscopy.

\subsection{VLT/FORS2 Spectroscopy}

Between 2011 July 22 and 2011 August 20 we obtained spectra for four of the six
that remained undetected in $I_N$ images. We used the FOcal Reducer/low
dispersion Spectrograph 2 \citep[FORS2;][]{app98} on the 8.2\,m Very Large
Telescope (VLT) Antu. The candidates were observed through the 600z
holographic grism with a 1\farcs3 wide longslit. The pixels were 2$\times$2
binned to decrease the readout time and noise, giving a spatial scale
0\farcs25 pixel$^{-1}$ and a dispersion of 1.62\,\AA\,pixel$^{-1}$. Between
exposures the pointing of the telescope was shifted by $\sim$10\arcsec\ along
the slit for more accurate sky subtraction. The position angle of the slit was
set to include a nearby star that could provide the trace of the spectrum.

\begin{figure}
\epsscale{1.20}
\plotone{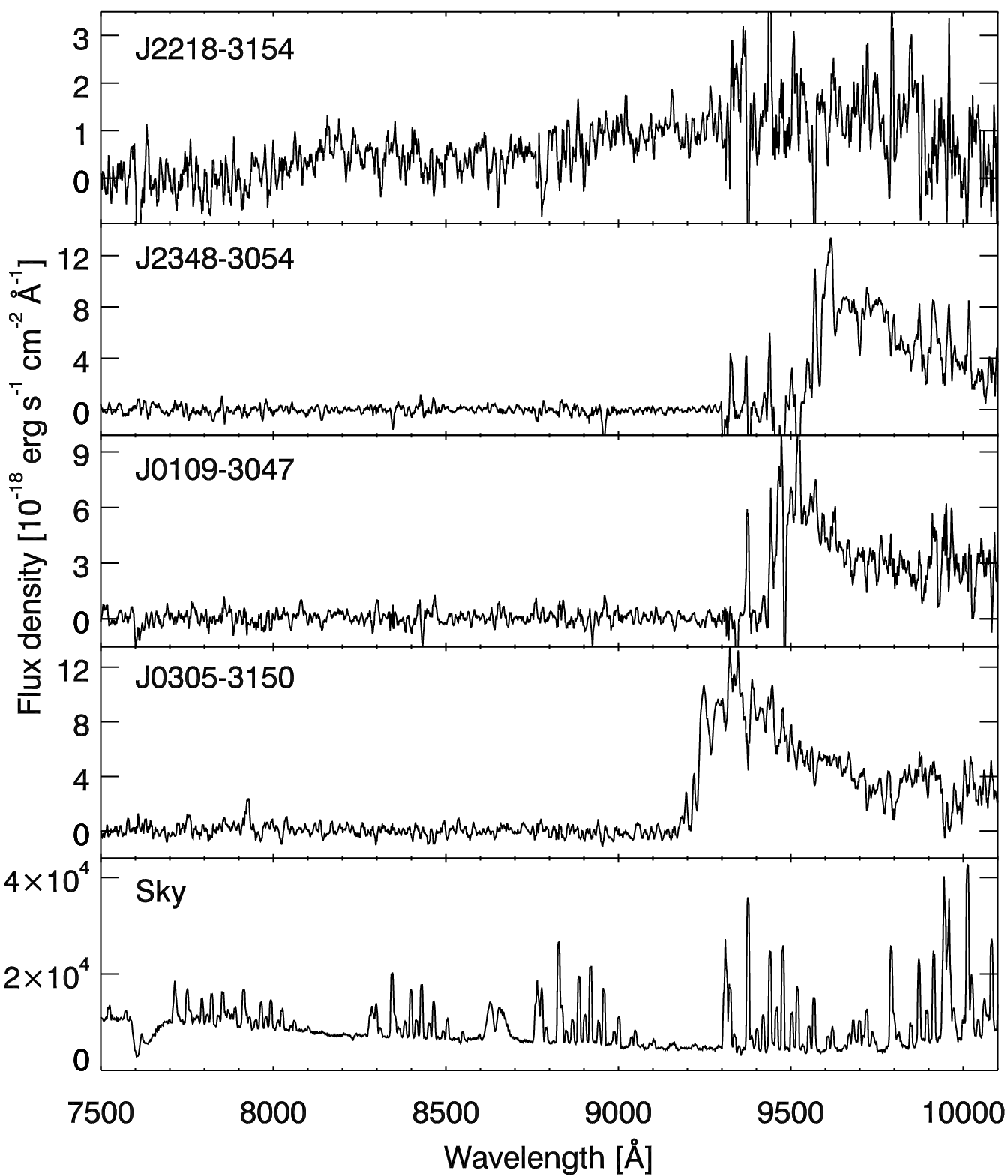}
\caption{VLT/FORS2 spectra of four high-redshift quasar candidates that
  remained undetected in the $I_N$ images. Below the four objects a sky
  spectrum is plotted. The spectra are boxcar averaged over five pixels. The
  spectra were not corrected for slit losses. Three of the targets
  (J2348--3054, J0109--3047, and J0305--3150) show a continuum decrement above
  $\sim$9200\,\AA, which is characteristic for quasars at $z\gtrsim6.5$. The
  fourth target (J2218--3154) does not show a break in the continuum and is
  probably an L dwarf. Identifying the peak of the quasar spectra as
  Ly$\alpha$ (1216\,\AA), the estimated redshifts for the quasars are
  $z\sim6.9$ (J2348--3054), $z\sim6.8$ (J0109--3047), and $z\sim6.6$
  (J0305--3150). \label{forsspectra}}
\end{figure}

On 2011 July 22, 23, and 29 candidate VIKING J234833.34--305410.0 (hereafter
J2348--3054) was observed for a total of 6750\,s under varying conditions. On
July 22 and 23 the seeing varied between 1\farcs5 and 2\farcs5 with clear
sky. On July 29 the seeing was 1\farcs1--1\farcs3 while there were some thin
clouds. Candidate VIKING J010953.13--304726.3 (hereafter J0109--3040) was
observed for 2280\,s on 2011 July 22. Conditions were clear with a seeing
around 1\farcs2. Six exposures totaling 4500\,s were taken of candidate VIKING
J030516.92--315056.0 (hereafter J0305--3150) on 2011 August 13 and 20. The
seeing was around 0\farcs7, and there were some thin clouds. Finally candidate
VIKING J221857.36--315430.0 (hereafter J2218--3153) was observed on 2011
August 20 for 1350\,s through thin clouds and with a seeing of 0\farcs6.

The data reduction, which included bias subtraction, flat fielding using lamp
flats and sky subtraction, was done using custom written routines in IDL. For
the wavelength calibration exposures of He, HgCd, and Ne arc lamps were
obtained. The wavelength calibration was checked using sky emission lines. The
typical rms of the wavelength calibration was better than 0.6\,\AA. The
spectra cover roughly the wavelength range 7100\,\AA--10300\,\AA. Observations
with a 5\arcsec\ slit of the spectrophotometric standard star LTT1020
\citep{ham92,ham94} were used for the flux calibration. The flux calibration
was found to be highly uncertain above 10100\,\AA\ due to a lack of flux
density values for LTT1020 in the literature above this value. We therefore
ignore the spectra above this wavelength.
The FORS2 spectra are shown in Figure \ref{forsspectra}.

\subsection{VLT/X-Shooter Observations}
\label{xshooter}

To increase the wavelength coverage of our spectroscopy to the
near-infrared, we observed two of our spectroscopic targets,
J2348--3054 and J0109--3040, with the medium resolution spectrograph
X-Shooter \citep{ver11} on the Cassegrain focus of the 8.2\,m VLT
Kueyen (UT2) on 2011 August 19--21 and on 2011 November 24. X-Shooter
consists of three spectrographs, each covering a different wavelength
region. The three arms and wavelength coverage of the instrument are
as follows: UVB (3000\,\AA--5595\,\AA), VIS (5595\,\AA--10240\,\AA)
and NIR (10240\,\AA--24800\,\AA). Given the extreme red colors of our
targets, we ignored the UVB arm. The slit length in the VIS and NIR
arm was 11\arcsec, and the pixel scales are 0\farcs16\,pixel$^{-1}$
and 0\farcs21\,pixel$^{-1}$, respectively. The slit width used for the
observations ranged between 0\farcs9--1\farcs5 in the VIS arm and
between 0\farcs9--1\farcs2 in the NIR arm, with the width adjusted to
counter periods of bad seeing. The resulting resolution varied between
$R=5400-8800$ in the VIS arm and between $R=4000-5000$ in the NIR
arm. Observing conditions on 2011 August 19 were variable with the
seeing in the optical ranging from 0\farcs9 to $\sim$1\farcs3 and some
thin clouds. For flux calibration, observations of the
spectrophotometric standard stars LTT7987 and Feige110 were obtained
\citep{ham94,ver10}.

We observed J0109--3040 for 14,400\,s in August and for 7200\,s in
November. J2348--3054 was observed in August only for 8783\,s. We made
use of the X-Shooter pipeline version 1.3.7 \citep{mod10} to reduce
the data to two-dimensional rectified wavelength-calibrated
spectra. The extraction of the spectra and the flux calibration were
performed with custom-made programs in IDL. We scaled the
one-dimensional spectra to match the observed infrared colors from
VIKING (listed in Table \ref{qsocolors}). We do note that the $Y-J$
color of the quasars derived from the X-Shooter spectra is 0.2\,mag
redder as compared to the catalog colors (Table \ref{qsocolors}). A
possible reason is that the light that the $Y$-band is measuring ends
up partly in the VIS arm and in the NIR arm. This causes additional
uncertainties in the flux calibration around 1\,$\mu$m. In the top two
panels of Figure \ref{irspecs} we show the reduced X-Shooter spectra.

\subsection{Magellan/FIRE Observations}
\label{fire}

\begin{deluxetable}{lccc}[t]
\tablecaption{Photometric Properties and Derived Parameters of the Three
    $z>6.5$ Quasars \label{qsocolors}}
\tablewidth{0pt} 
\tablehead{ \colhead{} & \colhead{J2348--3054} & \colhead{J0109--3047} &
  \colhead{J0305--3150}}
\startdata
RA (J2000) & 23$^h$48$^m$33$^s\!\!$.34 & 01$^h$09$^m$53$^s\!\!$.13 & 
03$^h$05$^m$16$^s\!\!$.92 \\
DEC (J2000) & --30$^\circ$54\arcmin10\farcs0 &
--30$^\circ$47\arcmin26\farcs3 &  --31$^\circ$50\arcmin56\farcs0 \\
$Z_{N,\mathrm{AB}}$ & 22.97$\pm$0.13 & 22.93$\pm$0.16 & 22.12$\pm$0.07 \\
$Z_{\mathrm{AB}}$ & $>$23.42 & $>$22.98 & 23.10$\pm$0.35 \\
$Y_{\mathrm{AB}}$ & 21.11$\pm$0.08 & 21.22$\pm$0.14 & 20.89$\pm$0.07 \\
$J_{\mathrm{AB}}$ & 21.14$\pm$0.08 & 21.27$\pm$0.16 & 20.68$\pm$0.07 \\
$H_{\mathrm{AB}}$ & 20.80$\pm$0.13 & 21.03$\pm$0.17 & 20.72$\pm$0.11 \\
$K_{\mathrm{s,AB}}$ & 20.51$\pm$0.10 & -- & 20.27$\pm$0.09 \\
$z_\mathrm{MgII}$ & $6.886_{-0.008}^{+0.009}$ & $6.745_{-0.009}^{+0.010}$ & 
$6.604\pm0.008$ \\
$M_\mathrm{1450}$ & --25.72$\pm$0.14 & --25.52$\pm$0.15 & --25.96$\pm$0.06 \\
\mbh\ (\msun) & $(2.1$$\pm$$0.5)$$\times$$10^9$ & $(1.5$$\pm$$0.4)$$\times$$10^9$ &
$(1.0$$\pm$$0.1)$$\times$$10^9$\\ 
\enddata
\tablecomments{Magnitude limits are 3$\sigma$.}
\end{deluxetable}

We observed J0305--3150 using the Folded-Port Infrared Echellette
\citep[FIRE;][]{sim08,sim10} on the Magellan Baade telescope on four
separate observing runs between 2011 November and 2012 January. FIRE
produces spectra with $R=6000$ ($\Delta v = 50$\,km\,s$^{-1}$) over
the full $0.82-2.51\mu$m wavelength range.  The data were taken in
generally clear conditions with a seeing of 0\farcs8 FWHM or better,
with a total integration time of 26,400\,s.

We reduced each echelle frame using a customized IDL suite (FIREHOSE)
designed for FIRE data. The pipeline produces an order-by-order
two-dimensional spectral model of the sky using the principles
outlined in \citet{kel03} and subtracts this from the flattened data
frames. Wavelength solutions are obtained using tabulated values of
the OH sky emission lines imprinted on the same frames. We
interspersed observations of A0V stars for telluric correction, which
was performed using the algorithms of \citet{vac03} as implemented in
the Spextool software package \citep{cus04}. Finally, the corrected
and flux calibrated orders for all exposures were combined into a
single one-dimensional spectrum, with each individual sampling
weighted by its squared S/N for an optimum S/N composite (Figure
\ref{irspecs}).

\section{RESULTS}
\label{results}

\begin{figure*}
\epsscale{.90}
\plotone{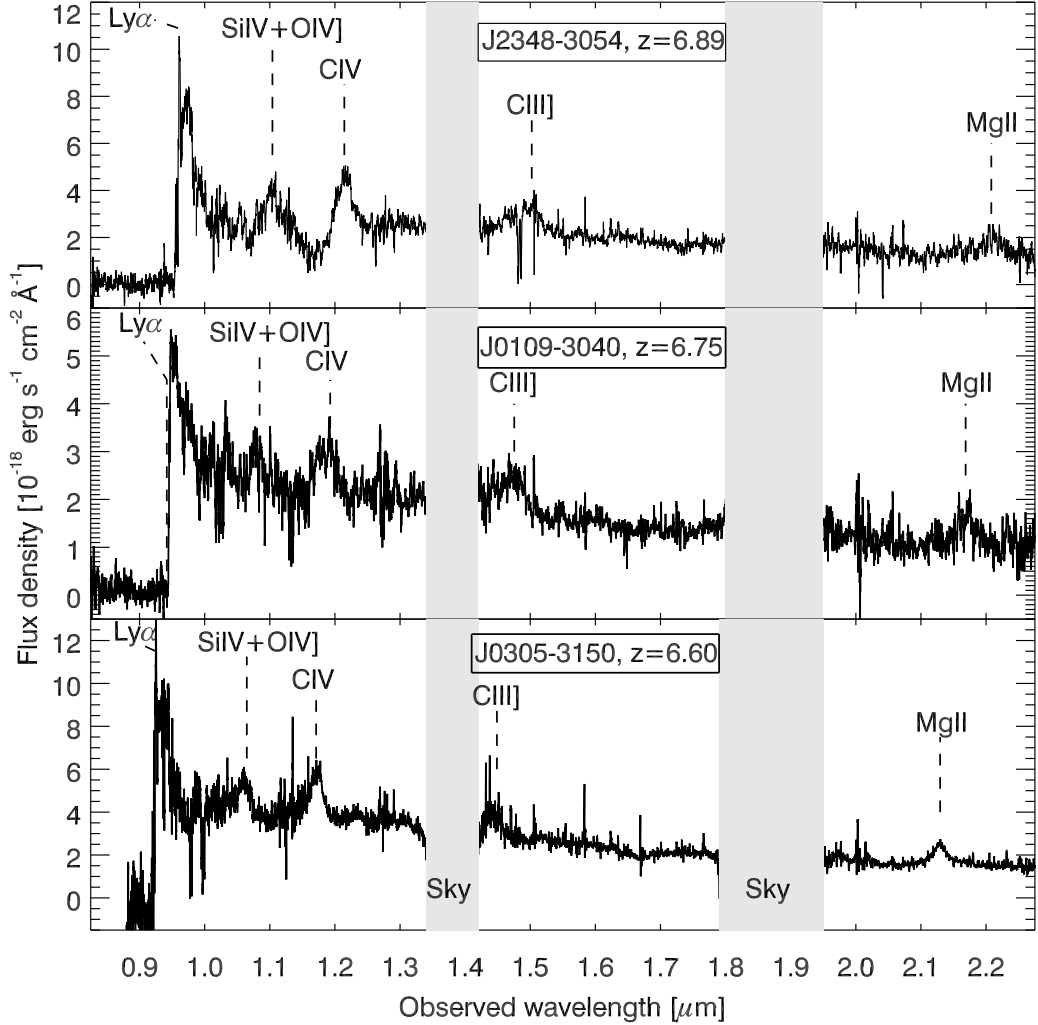}
\caption{X-Shooter (top two panels) and FIRE (bottom panel) medium
  resolution, infrared spectra of the three new $z>6.5$ quasars. The
  spectra are smoothed over 11\,pixels to better show the broad
  emission lines. Various strong emission lines in the spectra are
  marked. \label{irspecs}}
\end{figure*}

We obtained $I_N$ imaging of 45 high redshift quasar candidates with the NTT
and $Z_N$ imaging for more promising candidates. Based on the colors (Figure
\ref{nttccplot}), 39 objects could be rejected as quasar candidates after the
NTT imaging as they are likely to be late M or early L dwarfs. This is much as
expected, since these objects are numerous and a small fraction of them are
expected to scatter into our selection region due to random photometric
errors. Only six quasar candidates were undetected in the $I_N$ images (Figure
\ref{nttccplot} and Table \ref{efoscobs}) and could not immediately be
identified as a foreground object. Five of these satisfied the conservative
color criteria, while the sixth source, J2218--3153, was a filler object with
a (S/N)$_Y<7$. Four candidates, including J2218--3153, were spectroscopically
observed with FORS2 (Figure \ref{forsspectra}).

The spectrum of J2218--3153 shows a smoothly rising spectrum and the source is
most likely an early L dwarf. The FORS2 spectra of the three other objects,
J2348--3054, J0109--3047 and J0305--3150, show a strong continuum decrement
between 9200\,\AA\ and 9600\,\AA, and were identified as quasars at
$z=6.6-6.9$. For these three sources, we obtained near-infrared spectroscopy
(Figure \ref{irspecs}) to more accurately determine the redshift and estimate
the properties of the quasars. In Section \ref{method} we describe how we
derive the various quasar properties. These results are discussed in Section
\ref{properties} and summarized in Table \ref{qsocolors}.

Based on the available photometry, the two remaining candidates,
J115439.84+014153.3 and J115642.26--000758.1, could either be high
redshift quasars or late-type stars (M, L, or T dwarfs). Spectroscopic
observations of these sources are needed to determine their nature.

\subsection{$M_{1450}$, Redshift, and Black Hole Mass Estimates}
\label{method}

$M_{1450}$. The absolute magnitude of the quasars at a rest-frame wavelength of
1450\,\AA\ ($M_{1450}$) can be derived by measuring the flux density
in the near-infrared spectra at the redshifted wavelength
1450\,$\times$\,$(1+z)$\,\AA. The redshifts of the quasars of
$z=6.6-6.9$ shift the rest-frame wavelength of 1450\,\AA\ to an
observed wavelength of 1.10--1.15\,\mum, which is roughly between the
$Y$- and $J$-band. Because the $Y-J$ colors measured from the
X-Shooter spectra do not agree with the quasar colors from the
catalogs (see Sections \ref{xshooter}), we scale the spectra of
J2348--3054 and J0109--3040 to either the catalog $Y$ or $J$ magnitude
and compute $M_{1450}$ twice from the flux density that we measure in the
spectra at the appropriate wavelength. The absolute magnitudes of
these two quasars listed in Table \ref{qsocolors} are the average of
the two values, and the uncertainties are the standard deviation of the
mean. This way the uncertainty in the absolute magnitude includes the
issues with the flux calibration. For J0305--3150 the $Y-J$ color
measured from the FIRE spectrum was consistent with the one from the
catalog, and we could measure the flux density at 1450\,\AA\ in the
rest-frame directly in the spectrum.

{\it Black Hole Mass.} In \citet{der13} the infrared spectra of the quasars were fitted to estimate
the black hole mass and the redshift of the \ion{Mg}{2}\,$\lambda$2799
line. For the details of the fitting we refer to \citet{der13}. Here we will
give a short summary of the fitting procedures. \citet{der13} modeled the
quasar continuum as the sum of the power-law emission from the active galactic
nucleus, the Balmer continuum emission \citep{gra82} and the \ion{Fe}{2} and
\ion{Fe}{3} emission line forest \citep{ves01}. The probability of the
continuum model parameters was estimated using a grid-based approach. To fit
the \ion{Mg}{2} line, \citet{der13} sampled the obtained continuum probability
distribution $N$ times by using a Monte Carlo rejection method. For each
sample, the corresponding continuum was subtracted from the observed spectrum
and the emission line was fitted, assuming a Gaussian profile, using a
$\chi^2$ minimization routine.

The resulting distribution of black hole masses was estimated by combining the
$N$ continuum luminosities at $\lambda_{\mathrm{rest}}=3000$\,\AA\ with the
FWHM measured for the \ion{Mg}{2} line:
$M_\mathrm{BH}=10^{6.86}\,\mathrm{(FWHM(MgII)/1000)}^2\,(\lambda\,L_\mathrm{3000\AA}/10^{44})^{0.5}$
\citep{ves09} where $\mathrm{FWHM(MgII)}$ is the FWHM of the \ion{Mg}{2} line
in \kms\ and $\lambda\,L_\mathrm{3000\AA}$ is the nuclear continuum luminosity
at $\lambda_{\mathrm{rest}}=3000$\,\AA\ in erg\,s$^{-1}$\,\AA$^{-1}$.

The uncertainties concerning the black hole mass estimates listed in Section
\ref{properties} only include the uncertainties from the
continuum modeling and from the line fitting. Systematic uncertainties,
such as the scatter in the scaling relation of 0.55\,dex, are not
taken into account.

{\it Redshift.} We use the peak of the \ion{Mg}{2} emission line as measured by \citet{der13}
to obtain the redshift of the quasars. The redshifts derived this way have
been corrected for the velocity shift of $-97\pm269$\,\kms\ as found between
\ion{Mg}{2} and the systemic redshift of lower redshift quasars as traced by
the [\ion{O}{3}] $\lambda 5008$ emission line \citep{ric02b}. The uncertainty
in this shift is added in quadrature to the statistical redshift uncertainty
of the Gaussian fit of the emission line.

\section{PROPERTIES OF INDIVIDUAL QUASARS}
\label{properties}

\subsection{J2348--3054}

This source was selected with $Y-J=-0.03\pm0.11$ and $Z-Y>2.31$ (see Table
\ref{qsocolors}). With the NTT we observed the source for 5400\,s in $I_N$ and
for 3600\,s in $Z_N$. The source was not detected in the $I_N$ image, implying
$I_N>25.13$ and $I_N-Y>4.02$. Combined with a detection in the $Z_N$ image of
$Z_N=22.97\pm0.13$, we could identify this object as a probable high redshift
quasar with $6.5<z\lesssim7$ and rule out a foreground identification of the
source (Figure \ref{nttccplot}).

The discovery spectrum taken with VLT/FORS2 (Figure \ref{forsspectra}) shows a
source with continuum detected in the red part and a sharp break between
9500\,\AA\ and 9600\,\AA. We interpreted this feature as the onset of
absorption in the IGM shortward of the Ly$\alpha$ line in the quasar and
estimated a redshift of $z\sim6.9$. The VLT/X-Shooter spectrum (Figure
\ref{irspecs}) shows several strong emission lines, such as
\ion{Si}{4}\,$\lambda$1397\,+\,\ion{O}{4}]\,$\lambda$1402,
\ion{C}{4}\,$\lambda$1549, \ion{C}{3}]\,$\lambda$1909, and
\ion{Mg}{2}\,$\lambda$2799. A fit of the \ion{Mg}{2} line produced a more
accurate redshift of $z=6.886_{-0.008}^{+0.009}$. The continuum blueward of
the \ion{C}{4} emission line is partly absorbed, suggesting that this quasar
is a broad absorption line (BAL) quasar.

Extrapolating the continuum above the \ion{C}{4} absorption, we measure an
absolute magnitude at 1450\,\AA\ in the rest-frame of
$M_{1450}=-25.72\pm0.14$, which is approximately 1\,mag fainter than ULAS
J1120+0641 at $z=7.1$ \citep{mor11}.

The black hole mass that \citet{der13} derive from the width of the
\ion{Mg}{2} emission line and the strength of the continuum is
\mbh\,=\,$(2.1\pm0.5)\times10^9$\,\msun.

\subsection{J0109--3040}

In our catalogs, this source was selected with colors $Y-J=-0.05\pm0.21$
and $Z-Y>1.76$ (Table \ref{qsocolors}). We observed the source for 8100\,s
in $I_N$ and 3600\,s in $Z_N$ with the NTT. Similar to J2348--3054,
J0109--3040 remained undetected in the deep $I_N$ image, suggesting a very red
$I_N-Y > 4.05$. The $Z_N-J$ was slightly ($\sim$0.2\,mag) bluer then that
of J2348--3054, again ruling out a foreground interpretation of the
optical--infrared colors. The FORS2 spectrum revealed an object with a strong
continuum break around 9400\,\AA--9500\,\AA\ (Figure \ref{forsspectra}), and we
identified the source as a quasar at $z\sim6.8$.

The X-Shooter spectrum (Figure \ref{irspecs}) shows several strong
emission lines, including \ion{Si}{4}\,+\,\ion{O}{4}, \ion{C}{3}], and
\ion{Mg}{2}, from which we derive a redshift of
$z_\mathrm{MgII}=6.745_{-0.009}^{+0.010}$. The Ly$\alpha$ emission line is
completely absorbed. Despite the lack of Ly$\alpha$ emission, the
continuum at wavelengths shortward of \ion{C}{4} is not absorbed and,
unlike J2348--3054, J0109--3040 is not a BAL.

From the X-Shooter spectrum we can directly measure the flux density
at 1450\,\AA\ in the rest frame. We find that the quasar has
$M_{1450}=-25.52\pm0.15$. The uncertainty in the absolute magnitude
includes the moderate S/N of the $Y$ and $J$ VIKING magnitudes that
were used to derive the absolute flux calibration of the spectrum. The
absolute magnitude of $M_{1450}=-25.52$ makes it the faintest quasar
in our sample and 1.5--2.5\,mag fainter than the $z\sim6$ quasars
found in the main SDSS survey \citep{fan06b}. Although the quasar is
fainter in the continuum, the black hole mass of
\mbh\,=\,$(1.5\pm0.4)\times10^9$\,\msun\ is similar to that of
J2348--3054 \citep{der13}.

\subsection{J0305--3150}
\label{j0305m3150}

\begin{figure*}[t]
\epsscale{1.15}
\plottwo{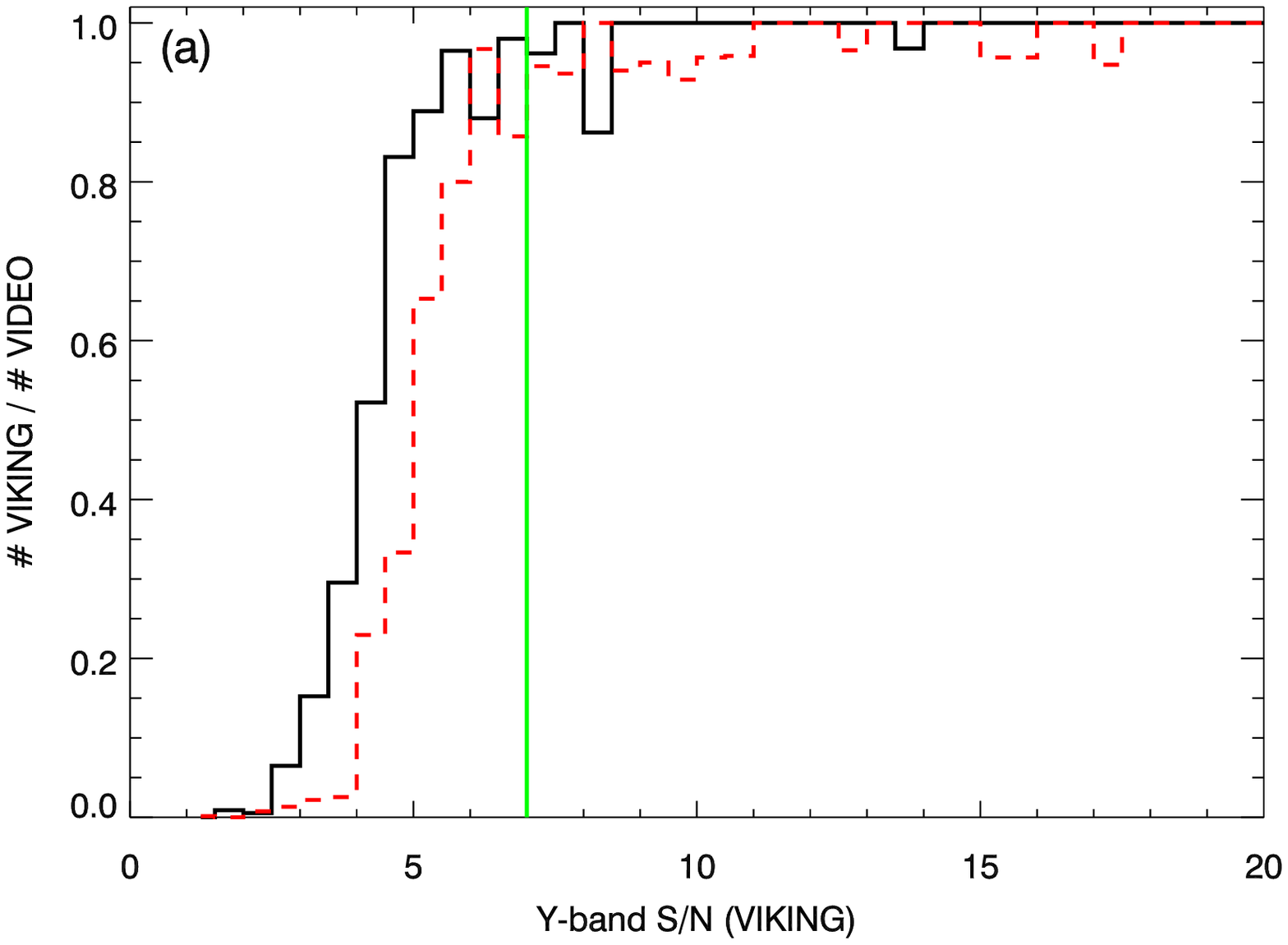}{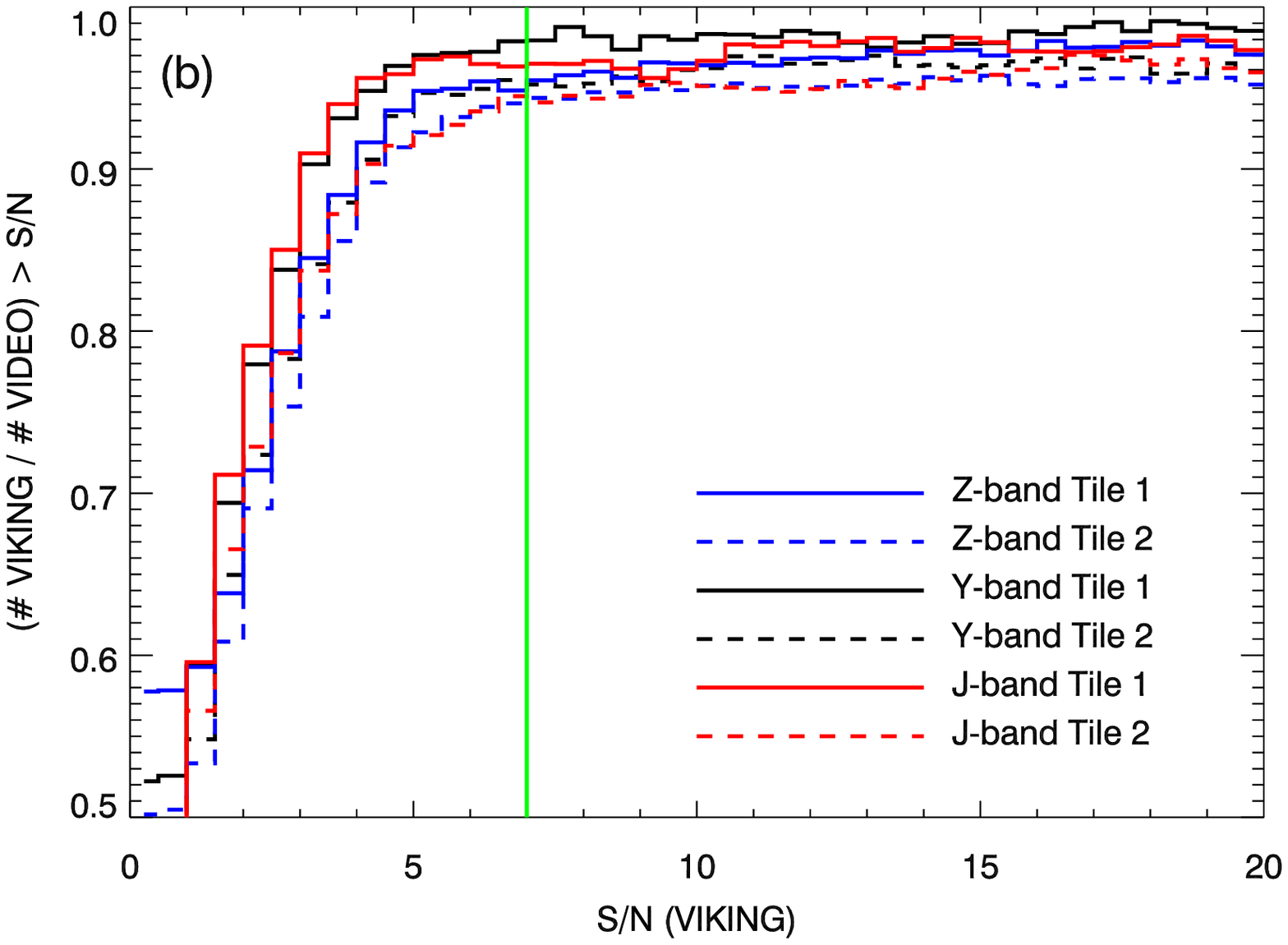}
\caption{(a) Completeness of the $Y$-band counts as a function of $Y$-band
  S/N. The fractions are derived by comparing the counts in the VIKING
  catalogs with the counts in the VIDEO catalog in similar magnitude
  ranges. The solid and dashed histograms represent two different VIKING tiles
  that (partly) overlap with the VIDEO data. The vertical solid line indicates
  the 7$\sigma$ limit we applied to the VIKING catalogs. (b) VIKING
  $Z$-, $Y$-, and $J$-band catalog completeness as a function of limiting S/N
  for the two tiles overlapping with the VIDEO survey. Down to 7$\sigma$
  the VIKING catalogs are 94\%--98\% complete. \label{sncompleteness}}
\end{figure*}

J0305--3150 was the only quasar in our sample that was detected in the VIKING
$Z$ image. The $Z-Y$ color of $Z-Y=2.21\pm0.36$ was red enough
to fulfill the conservative color selection. Over the three observing nights
with the NTT we exposed the source for a total of 3600\,s through the $I_N$
filter and for 2700\,s through $Z_N$. While the $I_N-Y$ color we measured was
similar to that of the other two quasars ($I_N-Y>4.08$; Table
\ref{qsocolors}), the $Z_N-J$ was 0.2--0.4\,mag bluer. This was an indication
that the redshift is lower than that of the other quasars. Of the three new
quasars, this object had the brightest (observed) magnitudes, being around
0.5\,mag brighter in $J$ than the other two quasars.

The FORS2 observations showed a quasar spectrum with a break around
$\sim$9200\,\AA, corresponding to a Ly$\alpha$ redshift of $z\sim6.6$. The
redshift was more accurately determined using the FIRE spectrum of the
quasar. The infrared spectrum showed the same emission lines as in the other
two VIKING quasars. A fit to the \ion{Mg}{2} emission line gives a redshift of
$z=6.604\pm0.008$. The absolute magnitude of this quasar is
$M_{1450}=-25.96\pm0.06$ making it the brightest of the new quasars, but still
$\sim$0.6\,mag fainter than ULAS J1120+0641. 

The wavelength coverage of the FIRE spectrum allowed \citet{der13} to estimate
the mass of the black hole. Based on the fit to the continuum and on the
Gaussian modeling of the \ion{Mg}{2} emission line, they derive a
\mbh\,=\,$(1.0\pm0.1)\times10^9$\,\msun.

\section{THE QUASAR LUMINOSITY FUNCTION AT $z\sim7$}
\label{qsolfsection}

In this section we will compare our discovery of three quasars at $z>6.5$ in
331.6\,\sqdeg\ that was covered by VIKING with the expected number of quasars
based on the quasar luminosity function at $z\sim6$. We first estimate the
completeness of the catalogs and the efficiency of the point source selection
(Sections \ref{catcomp} and \ref{pscomp}). We then test the efficiency and
completeness of our conservative and extended color selections (Section
\ref{colcomp}). In Section \ref{qsolf} we apply the completeness calculations
to the number density of VIKING quasars and compare those with the predictions
based on the $z=6$ quasar luminosity function extrapolated to $z>6.5$.

\subsection{Catalog Completeness}
\label{catcomp}

To assess how complete the VIKING catalogs are as a function of
limiting S/N, we need to compare the detected number of sources in the
VIKING catalogs with the expected number of sources. To determine the
number of objects that are expected in the VIKING images, we make use
of the first data release of the VISTA Deep Extragalactic Observations
(VIDEO) survey \citep{jar13}. The VIDEO data overlap with test data
from the VIKING survey in the Canada--France--Hawaii-Telescope-Legacy
Survey Deep-1 field. Since the VIDEO survey is
2.2--2.6\,mag deeper than VIKING in all five VISTA bands and highly
complete down to the limits of the VIKING survey \citep[see Table 3
in][]{jar13}, we can take the VIDEO catalogs to determine which
fraction of the sources is recovered in the VIKING catalogs. The VIDEO
area is partly covered by two different VIKING pointings, with the
total overlap roughly 1\,\sqdeg. The two VIKING tiles that overlap the
VIDEO region have similar characteristics (average limiting
magnitudes, seeing, etc.) as those of a typical VIKING tile.

Our calculation of the completeness of the VIKING catalogs as a function of
S/N, was done in the following way. First, we computed for a given S/N what
the corresponding magnitude range was. We then counted the number of objects
within that magnitude range in the VIDEO catalogs and compared that to the
number of sources with a match in the VIKING catalogs. In Figure
\ref{sncompleteness}(a) we show the fraction of objects in the VIDEO $Y$-band
catalog that are recovered in the VIKING catalog. The VIKING catalogs are
over 95\% complete down to a S/N\,$\sim 5$. The completeness as
function of limiting S/N is plotted in Figure \ref{sncompleteness}(b).  We
computed the number of VIDEO sources above the magnitude corresponding to the
VIKING S/N limit and counted the fraction of sources that had a match above
the S/N limit in the VIKING catalog. We have six independent measurements of
the completeness (two VIKING pointings with three filters each), and the
completeness above a S/N\,$>7$ was in the range 94.4\%--97.9\% with an average
of 95.7\%.

Based on the counts in the deeper VIDEO catalogs, we recover around
96\% of the objects in the VIKING catalogs above our S/N cut of 7. Therefore,
the catalog completeness correction is very small and will not have a strong
influence on the results presented in this Section.

\subsection{Point Source Completeness}
\label{pscomp}

\begin{figure*}
\epsscale{0.8}
\plotone{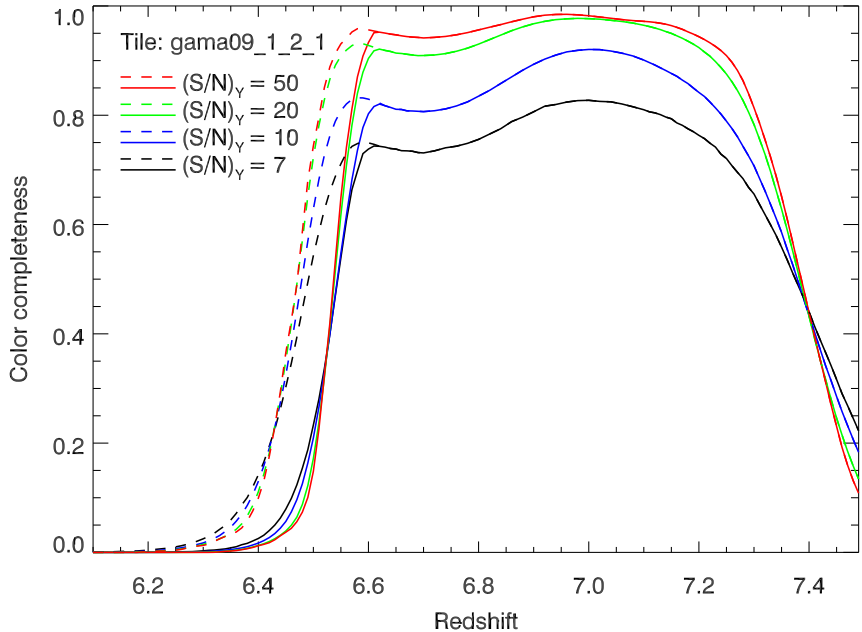}
\caption{Color completeness as a function of redshift for tile
  gama09\_1\_2\_1. The solid lines illustrate the completeness for the
  conservative color criteria, while the dashed lines represent the
  completeness when using the extended color criteria. The dashed lines
  overlap with the solid lines beyond $z\gtrsim6.6$. From top to bottom the
  different lines demonstrate the change in completeness as a function of
  decreasing (S/N)$_Y$, going from (S/N)$_Y=50$ to (S/N)$_Y=7$. The extended
  color criteria is more sensitive for quasars in the range $6.5 \lesssim z
  \lesssim 6.6$ as compared to the conservative color
  criteria. \label{tilecomp}}
\end{figure*}

To remove foreground galaxies from our quasar color selection, we only
selected point sources by forcing a pGalaxy\,$< 0.95$ (Section
\ref{candidateselection}). We tested the efficiency and completeness of this
criterion in two different ways. First, we matched the VIKING data in the
GAMA09 region with objects from the SDSS database \citep[Data
Release 8;][]{aih11}. In a second test we used the VIDEO data to select a
clean, photometric sample of stars in the region for which VIKING data exist.

In the GAMA09 region, there are 3653 objects that were classified as point
sources in the SDSS imaging catalogs and had a spectroscopic classification
``STAR''. In the same region, we selected 12618 objects that were classified
as galaxies in both the imaging and spectroscopic SDSS data. Using our adopted
point-source classifier, pGalaxy\,$< 0.95$, we selected 3397 of the SDSS
point sources and only 9 of the galaxies (0.07\%). Based on these numbers, we
estimate that our point-source completeness is 93.3\%, and our sample has a
purity of 99.7\%.

To obtain a clean sample of stars in the VIDEO data, we selected objects
for which the morphological information from the five VIDEO images ($Z$, $Y$,
$J$, $H$, and \ks) indicate that the source is stellar (i.e., the catalog
parameter ``mergedclass'' set to
--1 or --2). In addition, we required that the sources have optical--infrared
colors close to the stellar locus \citep[see e.g.\ Figure 1 in][]{fle12}. This
selection resulted in 2446 stellar sources in the VIDEO catalog that had a
match in the VIKING data. In the VIKING catalogs 2287 of the 2446 point
sources had pGalaxy\,$< 0.95$. This comparison implies that the VIKING
catalogs have a point-source completeness of 93.5\%. 

We find that the fraction of point sources misclassified in the VIKING
catalogs is small (6\%--7\%). Combined with the catalog completeness, we
estimate that we are missing 10.7\% of the point sources in our search.

\subsection{Color Selection Completeness}
\label{colcomp}

To estimate which fraction of quasars we are missing due to our color
criteria, we follow the approach by \citet{wil05} to clone SDSS spectra of
lower redshift quasars and shift them to higher redshifts. An alternative
method is to create artificial quasar spectra based on an empirical model for
the spectral energy distribution \citep[e.g.,][]{fan99,mcg13}. The main
assumption of both methods is that the rest-frame UV spectral properties of
quasars redward of the Ly$\alpha$ line do not evolve with redshift
\citep[e.g.,][]{kuh01,fan04}. Although the cloning of lower redshift quasar
spectra could suffer from incompleteness due to the SDSS quasar selection
function \citep[e.g.,][]{ric02a,wor11}, we decided to mimic the procedure of
\citet{wil05} as our goal is to compare our results to the $z\sim6$ quasar
luminosity function presented in \citet{wil10a} that was derived using the
same color completeness correction.

For the cloning we selected optical spectra of quasars from the fourth SDSS
Quasar Catalog \citep{sch07}. We needed to model colors of quasars in a redshift
interval covering at least $6.4 < z < 7.5$ in the filters $Z$, $Y$ and $J$
(Figure \ref{cc_zyj}). Given the wavelength range of the SDSS spectra ($3800
\lesssim \lambda \lesssim 9200$), we chose to use SDSS quasars in the redshift
range $3.4 < z < 3.5$ for the cloning. The SDSS archives the spectra of 428
quasars over this redshift interval; after culling for quality issues such
as missing data and incorrect redshift estimates, this number is reduced to
405 spectra, which can be artificially redshifted. The absolute magnitudes of
these quasars range from $M_{1450}=-24.8$ to $M_{1450}=-28.5$ with a median of
$M_{1450}=-25.8$, close to the average of our newly discovered quasars. 

\begin{figure}
\epsscale{1.25}
\plotone{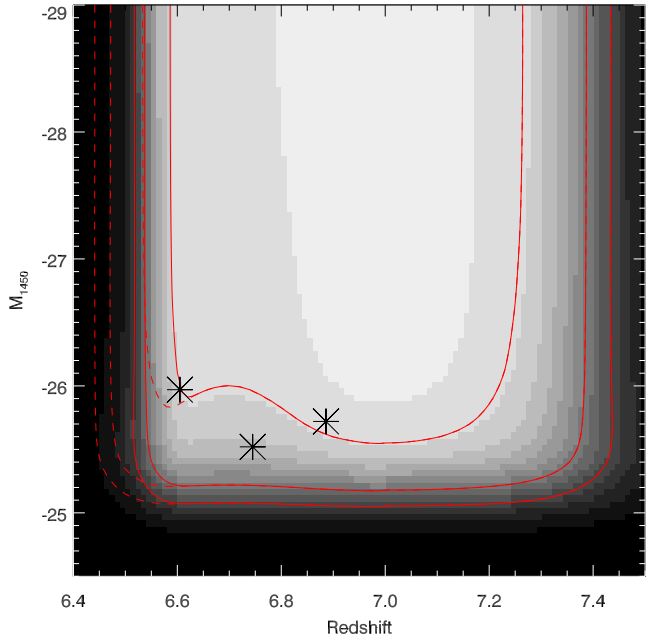}
\caption{Color completeness averaged over all tiles as a function of redshift
  and $M_{1450}$. The colors illustrate the completeness with the conservative
  color criteria going from 98.8\% (light gray) to 0\% (black). The red, solid
  (dashed) contours show the 90\%, 50\%, and 30\% completeness for the
  conservative (extended) color criteria. The large stars represent the
  redshift and absolute magnitude of our three new quasars. Our 50\% color
  completeness reaches an absolute magnitude of --25.2, and the 30\%
  completeness a slightly fainter absolute magnitude of
  --25.1. \label{avecomp}}
\end{figure}

The most significant difference between quasar spectra at $z\sim3.5$ and
$z\sim6.5$ is the spectral energy distribution blueward of the Ly$\alpha$
transition due to absorption by neutral hydrogen. Broadly speaking, as the
volume averaged \ion{H}{1} content of universe increases toward higher
redshifts, absorption of photons in the line of sight redshifted to the
wavelengths of Ly$\alpha$, Ly$\beta$, and higher order transitions becomes
stronger. When cloning high redshift quasar spectra from a lower redshift
sample this difference must be accounted for. We applied artificial absorption
to the SDSS spectra at $3.4<z<3.5$ following the procedure of
\citet{wil05}. In short, \citet{wil05} corrected for Ly$\alpha$ absorption
already present in the spectra of SDSS quasars using the IGM transmission as a
function of redshift as measured by \citet{son04}. The transmission was
sampled in large redshift intervals of $\Delta z=0.1$, which acted to smooth
out inhomogeneity of the IGM and reduce the difference between the average
optical depth per pixel and the effective optical depth. The spectra were then
further corrected for Ly$\beta$ absorption where the Ly$\beta$ optical depth
is given by $\tau_{\beta}=0.4 \times \tau_{\alpha}$ \citep{fan06b}. Higher
order absorption is negligibly small and was therefore left uncorrected.

Once these corrections were applied to the low-redshift spectra they were then
artificially redshifted and artificial absorption was added blueward of
Ly$\alpha$ and Ly$\beta$ according to the same relation from
\citet{son04}. Complete absorption of the spectra was applied blueward of the
Lyman limit (912\,\AA). The only difference between the procedure of
\citet{wil05} and our procedure is that we have updated the
transmission--redshift relation to reflect new measurements recently published
by \citet{bec13} for redshifts up to $z\lesssim5$. The transmission functions
from \citet{son04} and \citet{bec13} are in good agreement at $z\sim5$, so we
have not attempted to join the two functions.

Each of the artificial spectra was redshifted to $6.0<z<7.5$ in steps of
$\Delta z=0.01$ and multiplied by the transmission curves of the $Z$, $Y$, and
$J$ filters. The average fluxes in the filters were converted to observed
magnitudes. To take photometric errors into account in the color completeness,
we added random magnitude errors as a function of the S/N. Since each tile can
have different relative depths between the $Z$, $Y$, and $J$ images, we
computed the color completeness for each tile. An example of the resulting
color completeness as a function of redshift is shown in Figure
\ref{tilecomp}. The completeness in other tiles looks very similar. At high
S/N (i.e., negligible magnitude errors), the completeness for the conservative
color criteria rises steeply between $z=6.5$ and $z=6.6$ and remains above
$\sim$95\% up to $z=7.2$. At higher redshifts the completeness drops quickly
to 10\% at $z=7.5$. This is because at $z\gtrsim7.1$ the IGM absorption cuts
into the $Y$ bandpass, and the model quasars move redder in $Y-J$. At
$z\gtrsim7.4$ the average $Y-J$ color of the model quasars is $Y-J>0.5$,
outside our selection region. At lower S/N the peak completeness drops to
75\%--80\% for artificial quasars with (S/N)$_Y=7$, while the redshift
distribution broadens slightly. At the high redshift end, the completeness of
the extended color criteria is identical to the one of the conservative color
criteria. Due to the lower $Z-Y$ threshold in the extended color criteria, a
higher fraction of artificial quasars at $6.4<z<6.6$ are selected.

Around a completeness of 30\%--40\% the color completeness is almost independent
of S/N. We therefore take the redshift range where the completeness is above
30\% as the redshift range we probe with the VIKING data. For the conservative
color criteria, the redshift range is $6.51<z<7.44$, and for the extended
criteria it is $6.44<z<7.44$. The extended criteria are therefore sensitive
for quasars down to the redshift where the highest redshift quasars are found
in optical surveys \citep[e.g.,][]{wil07}. The surveyed area of 331.6\,\sqdeg\
and the probed redshift range of $\Delta z=0.93$ ($\Delta z=1.00$) for the
conservative (extended) color criteria translates into a survey volume of
2.7\,comoving Gpc$^3$ (2.9\,comoving Gpc$^3$).

To take the varying depth of the VIKING tiles (Figure \ref{areadepth}) into
account, we computed the color completeness for each tile as a function of
absolute magnitude $M_{1450}$ of the model quasars. In Figure \ref{avecomp} we
show the completeness averaged over all tiles as a function of both redshift
and absolute magnitude. The color completeness at $6.6<z<7.3$ drops to 50\%
(30\%) for quasars with an absolute magnitude of $M_{1450}=-25.2$ (--25.07).

We will now apply all the completeness corrections to our data
and compare the results with the expected number of quasars based on the
extrapolated luminosity function of $z=6$ quasars.

\subsection{The Quasar Luminosity Function at $z\gtrsim6.5$}
\label{qsolf}

\begin{figure*}[t]
\epsscale{0.8}
\plotone{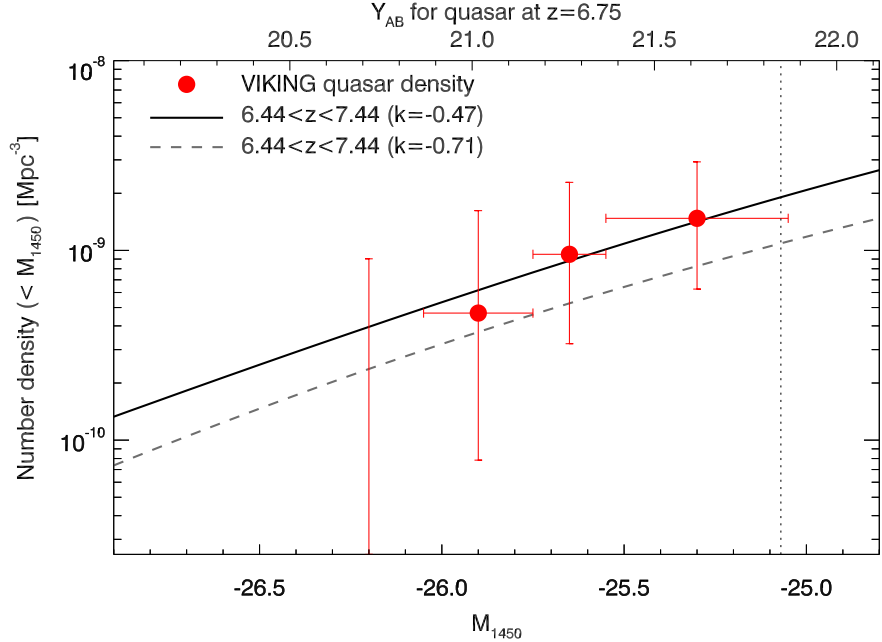}
\caption{Cumulative quasar luminosity function at $z\sim7$. The solid and
  dashed lines illustrate the predicted luminosity function in the redshift
  range $6.44<z<7.44$ based on the \citet{wil10a} $z=6$ quasar luminosity
  function and a decline in number density with redshift $\propto
  10^{-0.47\,z}$ and $\propto 10^{-0.71\,z}$, respectively. The number density
  of $6.44<z<7.44$ quasars implied by the new quasars has been computed in
  bins with a width of 0.1\,mag. All bins brighter than the one centered on
  $M_{1450}=-26.0$ only have an upper limit to the number density. Bins having
  the same number density were grouped together in a single data point with a
  (horizontal) error bar. For models to predict (within 1$\sigma$) the same
  number of quasars as we observe, they should fall in the region that is
  outlined by the vertical error bars. The vertical dotted line denote the
  absolute magnitude above which our color completeness is
  $>$30\%. \label{lumfunct}}
\end{figure*}

The quasar luminosity function, $\Phi(M,z$), is often parameterized by a
double power-law \citep[e.g.][]{boy00,fan01}:

\begin{equation}
  \Phi(M,z)\,=\,\frac{10^{k(z-6)}\Phi(M^*,\,z=6)}{10^{0.4(\alpha+1)(M-M^*)}\,+\,10^{0.4(\beta+1)(M-M^*)}},
\end{equation}

\noindent
where $\alpha$ is the faint end slope, $\beta$ the bright end slope,
$M$ the absolute magnitude, $M^*$ the break magnitude and
$\Phi(M^*,\,z=6)$ the space density at $z=6$ of quasars with a
magnitude $M^*$. The factor $10^{kz}$ accounts for the observed
decline with redshift in the space density of quasars between $z=3$
and $z>5$ \citep[e.g.,][]{sch95}.

\citet{wil10a} compiled a sample of 40 quasars at $z\sim6$ from
various optical surveys to derive the $z\sim6$ quasar luminosity
function. Fixing the faint end slope to $\alpha=-1.5$, based on the
quasar luminosity function at lower redshift \citep[e.g.][]{cro04},
\citet{wil10a} determined the following best-fit parameters for the
quasar luminosity function at $z=6$:
$\Phi(M^*_{1450})=1.14\times10^{-8}$\,Mpc$^{-3}$\,mag$^{-1}$,
$M^*_{1450}=-25.13$ and $\beta=-2.81$. A value of $k=-0.47$ was
adopted by \citet{wil10a} which fits the evolution of the (bright end
of the) luminosity function over the redshift range $3<z<6$
\citep{fan01}. Based on a study of $z\sim5$ quasars, \citet{mcg13}
derived a steeper decline in the space density of quasars with
$M_{1450} < -25.5$ between $z\sim4.9$ and $z=6$ of a factor $5.1$ per
unit redshift ($k=-0.71$).

If we assume that the quasar luminosity function evolves with constant
parameters $M^*_{1450}$, $\beta$ and $\alpha$ and pure density
evolution $\propto$$10^{kz}$, then we can extrapolate the $z=6$
luminosity function to $z>6.4$ and derive the expected number of
high-redshift quasars in the VIKING survey. Assuming $k=-0.47$ we
expect 0.0124 quasar\,deg$^{-2}$ down to $M_{1450}=-25.2$ in the
redshift range $6.51<z<7.44$, while the extrapolated luminosity
function predicts 0.0139 quasar\,deg$^{-2}$ in the redshift range
$6.44<z<7.44$. In an area of 331.6\,\sqdeg\ we therefore expect on
average $\sim$4.1 quasars between $6.51<z<7.44$ and $\sim$4.6 quasars
at $6.44<z<7.44$ if we ignore any incompleteness. The larger volume
probed by the extended color criteria and the higher predicted space
density of quasars at the lower end of our redshift range results in
an increase of 13\% in the expected number of quasars. With a sharper
decline in the normalization of $k=-0.71$, the expected numbers are a
factor of $\sim$1.75 smaller.

In Figure \ref{lumfunct} we show the cumulative luminosity function normalized
over our surveyed volume and the completeness-corrected space
density of $z\sim7$ quasars derived from our newly discovered VIKING quasars. 
Although the follow-up of all quasar candidates has not been
completed, the observed density of $z>6.44$ quasars is already fully
consistent with the one predicted from the extrapolated $z=6$ luminosity
function. A steeper decline in space density at $z>5$ as obtained by
\citet{mcg13} underpredicts the density of quasars at $z\sim7$ as found by our
study. If we take the various sources of incompleteness into account, then
with $k=-0.47$ we expect to find $\sim$3.1 $z>6.44$ quasars in our volume down
to $M_{1450}=-25.2$. Assuming $k=-0.71$ we expect 1.8 quasars in the same
volume. Therefore, given the large (Poisson) error on these estimates, we
cannot rule out the steeper decline.

To estimate for which values of $k$ we would expect to find three quasars
in our survey volume, we extrapolated the $z=6$ luminosity function to
$6.44<z<7.44$ with $k$ ranging from $-$3 to 0 in steps of 0.01. For
each value of $k$, we integrated the luminosity function down to
$M_{1450}=-25.2$ and, including the various sources of incompleteness,
computed the expected number of quasars in our probed volume. The best-fit 
value of $k$ for which the expected number is three
quasars and the 1$\sigma$ confidence interval is
$k=-0.49^{+0.28}_{-0.74}$. The best-fit value of $k=-0.49$ is close to
the $k=-0.47$ \citet{wil10a} adopts. The best-fit value obtained by
\citet{mcg13}, $k=-0.71$, is nonetheless well within the range of $k$
that fit our results.

\section{SUMMARY AND DISCUSSION}
\label{discussion}

We used the VIKING near-infrared imaging survey to search for quasars above a
redshift $z\gtrsim6.5$. We analyzed 332\,\sqdeg\ of VIKING data down to a
median 7$\sigma$ $Y$-band depth of $21.5$. We employed two sets of color
criteria to select potential quasars at high redshift. Our conservative color
criteria is sensitive to quasars in the redshift range $6.51<z<7.44$ and with
these criteria we selected 43 quasar candidates. Our extended color criteria
probed an increased redshift range of $6.44<z<7.44$ and selected an additional
42 objects. After we obtained optical imaging of 45 quasar candidates with the
NTT, 6 objects remained having colors consistent with high redshift
quasars. Spectroscopic follow-up with VLT/FORS2 of four of these sources
identified three previously unknown quasars with redshifts of $z=6.6-6.9$. The
fourth object was a foreground star. The newly discovered quasars have the
following characteristics:

\begin{enumerate}

\item The absolute magnitudes of the new quasars are in the range of
  $-25.5<M_{1450}<-26.0$, which is about 1--2\,mag fainter than the $z\sim6$
  quasars discovered in the SDSS main survey \citep{fan06b} and roughly similar
  to the quasars at $z\sim6$ found in the deep stripe of the SDSS southern
  survey \citep{jia08}. The only other $z\sim7$ quasar published so far,
  J1120+0641 at $z=7.1$, has an absolute magnitude of $-26.6$ \citep{mor11},
  0.6--1.1\,mag brighter than the VIKING quasars presented in this article.

\item Near-infrared spectroscopy revealed that the new VIKING quasars exhibit
  emission lines commonly seen in lower redshift quasars, including
  \ion{Si}{4}\,$\lambda$1397\,+\,\ion{O}{4}]\,$\lambda$1402,
  \ion{C}{4}\,$\lambda$1549, \ion{C}{3}]\,$\lambda$1909, and
  \ion{Mg}{2}\,$\lambda$2799. Furthermore, J2348--3054 shows continuum
  absorption blueward of the \ion{C}{4} emission line and could be a BAL
  quasar. 

\item Using the infrared spectra, \citet{der13} determined the mass of the
  central black hole powering the quasars. By fitting a Gaussian to the
  \ion{Mg}{2} line and measuring the continuum luminosity, they estimate
  central black hole masses of $1-2\times10^9$\,\msun. These black hole masses
  are very similar to those powering quasars at $z\sim6$ that were discovered
  in the SDSS main survey \citep[e.g.,][]{jia07,kur07,der11}, but generally
  more massive than those of the $z\sim6$ quasars discovered in the CFHQS
  \citep{wil10b}. Ignoring any incompleteness in our quasar search we can set
  a lower limit to the density of supermassive black holes in the early
  universe. The discovery of three black holes with masses
  $\sim1-2\times10^9$\,\msun\ in the 2.9\,comoving Gpc$^3$ probed in this
  study implies a minimum density of $\rho(M_\mathrm{BH}>10^9\,M_\odot) >
  1.1\times10^{-9}$\,\cmpc.

\end{enumerate}

To derive the space density of quasars in the volume probed by the VIKING
data, we computed the efficiency of our selection criteria in order to correct
for the incompletenesses in our quasar search. The volume density of quasars
at $6.44<z<7.44$ that we derive agrees well with the extrapolated $z=6$ quasar
luminosity function. This includes a decline in the space density of quasars
of a factor 3 per unit redshift, which was previously obtained by measuring the
density of bright quasars between $3<z<6$. Our results do not support a
steeper decline in the space density of quasars with redshift.

These findings imply that we should discover a significant number of quasars
beyond $z\gtrsim6.5$ in the remainder of the VIKING area. The total area that
will be covered by VIKING will be 1500\,\sqdeg. If we ignore any
incompleteness, we expect 18--19 quasars between $6.51<z<7.44$ and $\sim$21
quasars at $6.44<z<7.44$ in the VIKING area down to $M_{1450} < -25.2$. With a
larger number of quasars we will be able to constrain the evolution of the
quasar luminosity function at $z\sim7$.

These new quasars are prime targets for various follow-up studies. In future
articles we will study the emission lines that can be seen in the near-infrared
spectra and examine the implications on the chemical enrichment near the black
holes. The optical spectra will be used to derive constraints on the column
density of neutral hydrogen along the line of sight. Furthermore, the Southern
declination of the sources makes them ideal targets for ALMA observations to
study the galaxies that host these quasars.

\acknowledgments

We thank the referee for valuable comments and suggestions.
GDR is grateful to the National Science Foundation for support of this 
work through grant AST-1008882 to The Ohio State University.

Funding for SDSS-III has been provided by the Alfred P. Sloan
Foundation, the Participating Institutions, the National Science
Foundation, and the U.S. Department of Energy Office of Science. The
SDSS-III web site is http://www.sdss3.org/.

SDSS-III is managed by the Astrophysical Research Consortium for the
Participating Institutions of the SDSS-III Collaboration including the
University of Arizona, the Brazilian Participation Group, Brookhaven National
Laboratory, University of Cambridge, Carnegie Mellon University, University of
Florida, the French Participation Group, the German Participation Group,
Harvard University, the Instituto de Astrofisica de Canarias, the Michigan
State/Notre Dame/JINA Participation Group, Johns Hopkins University, Lawrence
Berkeley National Laboratory, Max Planck Institute for Astrophysics, Max
Planck Institute for Extraterrestrial Physics, New Mexico State University,
New York University, Ohio State University, Pennsylvania State University,
University of Portsmouth, Princeton University, the Spanish Participation
Group, University of Tokyo, University of Utah, Vanderbilt University,
University of Virginia, University of Washington, and Yale University.

{\it Facilities:} \facility{ESO:VISTA}, 
\facility{NTT(EFOSC2)},
\facility{VLT:Antu(FORS2)}, 
\facility{VLT:Kueyen(X-Shooter)},
\facility{Magellan:Baade(FIRE)}.


\begin{thebibliography}{66}
\expandafter\ifx\csname natexlab\endcsname\relax\def\natexlab#1{#1}\fi

\bibitem[{{Appenzeller} {et~al.}(1998){Appenzeller}, {Fricke}, {F{\"u}rtig},
    {G{\"a}ssler}, {H{\"a}fner}, {Harke}, {Hess}, {Hummel}, {J{\"u}rgens},
    {Kudritzki}, {Mantel}, {Meisl}, {Muschielok}, {Nicklas}, {Rupprecht},
    {Seifert}, {Stahl}, {Szeifert}, \& {Tarantik}}]{app98} 
{Appenzeller}, I., {Fricke}, K., {F{\"u}rtig}, W., et al. 1998, The
  Messenger, 94, 1

\bibitem[{{Aihara} {et~al.}(2011)}]{aih11} 
{Aihara}, H., Allende Prieto, C., An, D., et al. 2011, ApJS, 193, 29 

\bibitem[{{Arnaboldi} {et~al.}(2007){Arnaboldi}, {Neeser}, {Parker}, {Rosati},
  {Lombardi}, {Dietrich}, \& {Hummel}}]{arn07}
{Arnaboldi}, M., {Neeser}, M.~J., {Parker}, L.~C., et al. 2007, The Messenger,
127, 28

\bibitem[{{Becker} {et~al.}(2013){Becker}, {Hewett}, {Worseck}, \&
  {Prochaska}}]{bec13}
{Becker}, G.~D., {Hewett}, P.~C., {Worseck}, G., \& {Prochaska}, J.~X. 2013,
  MNRAS, 430, 2067

\bibitem[{{Becker} {et~al.}(2007){Becker}, {Rauch}, \& {Sargent}}]{bec07}
{Becker}, G.~D., {Rauch}, M., \& {Sargent}, W.~L.~W. 2007, ApJ, 662, 72

\bibitem[{{Bolton} {et~al.}(2011){Bolton}, {Haehnelt}, {Warren}, {Hewett},
  {Mortlock}, {Venemans}, {McMahon}, \& {Simpson}}]{bol11}
{Bolton}, J.~S., {Haehnelt}, M.~G., {Warren}, S.~J., et al.
  2011, MNRAS, L291

\bibitem[{{Bouwens} {et~al.}(2012){Bouwens}, {Illingworth}, {Oesch}, {Trenti},
  {Labb{\'e}}, {Franx}, {Stiavelli}, {Carollo}, {van Dokkum}, \&
  {Magee}}]{bou12}
{Bouwens}, R.~J., {Illingworth}, G.~D., {Oesch}, P.~A., et al. 2012, ApJL,
752, L5

\bibitem[{{Boyle} {et~al.}(2000){Boyle}, {Shanks}, {Croom}, {Smith}, {Miller},
  {Loaring}, \& {Heymans}}]{boy00}
{Boyle}, B.~J., {Shanks}, T., {Croom}, S.~M., et al. 2000, MNRAS, 317, 1014

\bibitem[{{Buzzoni} {et~al.}(1984){Buzzoni}, {Delabre}, {Dekker}, {Dodorico},
  {Enard}, {Focardi}, {Gustafsson}, {Nees}, {Paureau}, \& {Reiss}}]{buz84}
{Buzzoni}, B., {Delabre}, B., {Dekker}, H., et al. 1984, The Messenger, 38, 9

\bibitem[{{Croom} {et~al.}(2004){Croom}, {Smith}, {Boyle}, {Shanks}, {Miller},
  {Outram}, \& {Loaring}}]{cro04}
{Croom}, S.~M., {Smith}, R.~J., {Boyle}, B.~J., et al. 2004, MNRAS, 349, 1397

\bibitem[{{Cushing} {et~al.}(2004){Cushing}, {Vacca}, \& {Rayner}}]{cus04}
{Cushing}, M.~C., {Vacca}, W.~D., \& {Rayner}, J.~T. 2004, PASP, 116, 362

\bibitem[{{Dalton} {et~al.}(2006){Dalton}, {Caldwell}, {Ward}, {Whalley},
  {Woodhouse}, {Edeson}, {Clark}, {Beard}, {Gallie}, {Todd}, {Strachan},
  {Bezawada}, {Sutherland}, \& {Emerson}}]{dal06}
{Dalton}, G.~B., {Caldwell}, M., {Ward}, A.~K., et al. 2006, in Society of
Photo-Optical Instrumentation Engineers (SPIE) Conference Series, Vol. 6269,
Society of Photo-Optical Instrumentation Engineers (SPIE) Conference Series

\bibitem[{{De Rosa} {et~al.}(2011){De Rosa}, {Decarli}, {Walter}, {Fan},
  {Jiang}, {Kurk}, {Pasquali}, \& {Rix}}]{der11}
{De Rosa}, G., {Decarli}, R., {Walter}, F., et al. 2011, ApJ, 739, 56

\bibitem[{{De Rosa} {et~al.}(2013){De Rosa}, {Venemans}, {Decarli}, {Walter},
    {Fan}, {Jiang}, {Kurk}, {Pasquali}, \& {Rix}}]{der13} 
{De Rosa}, G., {Venemans}, B. P., {Decarli}, R., et al. 2013, ApJ, submitted (arXiv:1311.3260)

\bibitem[{{Emerson} {et~al.}(2006){Emerson}, {McPherson}, \&
  {Sutherland}}]{eme06}
{Emerson}, J., {McPherson}, A., \& {Sutherland}, W. 2006, The Messenger, 126,
  41

\bibitem[{{Fan}(1999)}]{fan99}
{Fan}, X. 1999, AJ, 117, 2528

\bibitem[{{Fan} {et~al.}(2004){Fan}, {Hennawi}, {Richards}, {Strauss},
  {Schneider}, {Donley}, {Young}, {Annis}, {Lin}, {Lampeitl}, {Lupton}, {Gunn},
  {Knapp}, {Brandt}, {Anderson}, {Bahcall}, {Brinkmann}, {Brunner}, {Fukugita},
  {Szalay}, {Szokoly}, \& {York}}]{fan04}
{Fan}, X., {Hennawi}, J.~F., {Richards}, G.~T., et al. 2004, AJ, 128, 515

\bibitem[{{Fan} {et~al.}(2001){Fan}, {Narayanan}, {Lupton}, {Strauss}, {Knapp},
  {Becker}, {White}, {Pentericci}, {Leggett}, {Haiman}, {Gunn}, {Ivezi{\'c}},
  {Schneider}, {Anderson}, {Brinkmann}, {Bahcall}, {Connolly}, {Csabai}, {Doi},
  {Fukugita}, {Geballe}, {Grebel}, {Harbeck}, {Hennessy}, {Lamb}, {Miknaitis},
  {Munn}, {Nichol}, {Okamura}, {Pier}, {Prada}, {Richards}, {Szalay}, \&
  {York}}]{fan01}
{Fan}, X., {Narayanan}, V.~K., {Lupton}, R.~H., et al. 2001, AJ, 122, 2833

\bibitem[{{Fan} {et~al.}(2006){Fan}, {Strauss}, {Becker}, {White}, {Gunn},
  {Knapp}, {Richards}, {Schneider}, {Brinkmann}, \& {Fukugita}}]{fan06b}
{Fan}, X., {Strauss}, M.~A., {Becker}, R.~H., et al. 2006, AJ, 132, 117

\bibitem[{{Findlay} {et~al.}(2012){Findlay}, {Sutherland}, {Venemans},
  {Reyl{\'e}}, {Robin}, {Bonfield}, {Bruce}, \& {Jarvis}}]{fin12}
{Findlay}, J.~R., {Sutherland}, W.~J., {Venemans}, B.~P., et al. 2012,
  MNRAS, 419, 3354

\bibitem[{{Fleuren} {et~al.}(2012){Fleuren}, {Sutherland}, {Dunne}, {Smith},
  {Maddox}, {Gonz{\'a}lez-Nuevo}, {Findlay}, {Auld}, {Baes}, {Bond},
  {Bonfield}, {Bourne}, {Cooray}, {Buttiglione}, {Cava}, {Dariush}, {De Zotti},
  {Driver}, {Dye}, {Eales}, {Fritz}, {Gunawardhana}, {Hopwood}, {Ibar},
  {Ivison}, {Jarvis}, {Kelvin}, {Lapi}, {Liske}, {Micha{\l}owski}, {Negrello},
  {Pascale}, {Pohlen}, {Prescott}, {Rigby}, {Robotham}, {Scott}, {Temi},
  {Thompson}, {Valiante}, \& {Werf}}]{fle12}
{Fleuren}, S., {Sutherland}, W., {Dunne}, L., et al. 2012, MNRAS, 423, 2407

\bibitem[{{Gnedin} \& {Fan}(2006)}]{gne06}
{Gnedin}, N.~Y. \& {Fan}, X. 2006, ApJ, 648, 1

\bibitem[{{Gonz{\'a}lez-Solares} {et~al.}(2008){Gonz{\'a}lez-Solares},
  {Walton}, {Greimel}, {Drew}, {Irwin}, {Sale}, {Andrews}, {Aungwerojwit},
  {Barlow}, {van den Besselaar}, {Corradi}, {G{\"a}nsicke}, {Groot}, {Hales},
  {Hopewell}, {Hu}, {Irwin}, {Knigge}, {Lagadec}, {Leisy}, {Lewis}, {Mampaso},
  {Matsuura}, {Moont}, {Morales-Rueda}, {Morris}, {Naylor}, {Parker}, {Prema},
  {Pyrzas}, {Rixon}, {Rodr{\'{\i}}guez-Gil}, {Roelofs}, {Sabin}, {Skillen},
  {Suso}, {Tata}, {Viironen}, {Vink}, {Witham}, {Wright}, {Zijlstra}, {Zurita},
  {Drake}, {Fabregat}, {Lennon}, {Lucas}, {Mart{\'{\i}}n}, {Phillipps},
  {Steeghs}, \& {Unruh}}]{gon08}
{Gonz{\'a}lez-Solares}, E.~A., {Walton}, N.~A., {Greimel}, R., et al. 2008,
MNRAS, 388, 89

\bibitem[{{Grandi}(1982)}]{gra82}
{Grandi}, S.~A. 1982, ApJ, 255, 25

\bibitem[{{Gunn} \& {Stryker}(1983)}]{gun83}
{Gunn}, J.~E. \& {Stryker}, L.~L. 1983, ApJS, 52, 121

\bibitem[{{Hamuy} {et~al.}(1994){Hamuy}, {Suntzeff}, {Heathcote}, {Walker},
  {Gigoux}, \& {Phillips}}]{ham94}
{Hamuy}, M., {Suntzeff}, N.~B., {Heathcote}, S.~R., et al. 1994, PASP, 106, 566

\bibitem[{{Hamuy} {et~al.}(1992){Hamuy}, {Walker}, {Suntzeff}, {Gigoux},
  {Heathcote}, \& {Phillips}}]{ham92}
{Hamuy}, M., {Walker}, A.~R., {Suntzeff}, N.~B., et al. 1992, PASP, 104, 533

\bibitem[{{Hewett} {et~al.}(2006){Hewett}, {Warren}, {Leggett}, \&
  {Hodgkin}}]{hew06}
{Hewett}, P.~C., {Warren}, S.~J., {Leggett}, S.~K., \& {Hodgkin}, S.~T. 2006,
  MNRAS, 367, 454

\bibitem[{{Irwin} {et~al.}(2004){Irwin}, {Lewis}, {Hodgkin}, {Bunclark},
  {Evans}, {McMahon}, {Emerson}, {Stewart}, \& {Beard}}]{irw04}
{Irwin}, M.~J., {Lewis}, J., {Hodgkin}, S., et al. 2004, in
  Society of Photo-Optical Instrumentation Engineers (SPIE) Conference Series,
  Vol. 5493, Society of Photo-Optical Instrumentation Engineers (SPIE)
  Conference Series, ed. P.~J. {Quinn} \& A.~{Bridger}, 411

\bibitem[{{Jarvis} {et~al.}(2013){Jarvis}, {Bonfield}, {Bruce}, {Geach},
  {McAlpine}, {McLure}, {Gonz{\'a}lez-Solares}, {Irwin}, {Lewis}, {Yoldas},
  {Andreon}, {Cross}, {Emerson}, {Dalton}, {Dunlop}, {Hodgkin}, {Le},
  {Karouzos}, {Meisenheimer}, {Oliver}, {Rawlings}, {Simpson}, {Smail},
  {Smith}, {Sullivan}, {Sutherland}, {White}, \& {Zwart}}]{jar13}
{Jarvis}, M.~J., {Bonfield}, D.~G., {Bruce}, V.~A., et al. 2013, MNRAS, 428,
1281

\bibitem[{{Jiang} {et~al.}(2008){Jiang}, {Fan}, {Annis}, {Becker}, {White},
  {Chiu}, {Lin}, {Lupton}, {Richards}, {Strauss}, {Jester}, \&
  {Schneider}}]{jia08}
{Jiang}, L., {Fan}, X., {Annis}, J., et al. 2008, AJ, 135, 1057

\bibitem[{{Jiang} {et~al.}(2009){Jiang}, {Fan}, {Bian}, {Annis}, {Chiu},
  {Jester}, {Lin}, {Lupton}, {Richards}, {Strauss}, {Malanushenko},
  {Malanushenko}, \& {Schneider}}]{jia09}
{Jiang}, L., {Fan}, X., {Bian}, F., et al. 2009, AJ, 138, 305

\bibitem[{{Jiang} {et~al.}(2007){Jiang}, {Fan}, {Vestergaard}, {Kurk},
  {Walter}, {Kelly}, \& {Strauss}}]{jia07}
{Jiang}, L., {Fan}, X., {Vestergaard}, M., et al. 2007, AJ, 134, 1150

\bibitem[{{Kelson}(2003)}]{kel03}
{Kelson}, D.~D. 2003, PASP, 115, 688

\bibitem[{{Komatsu} {et~al.}(2011){Komatsu}, {Smith}, {Dunkley}, {Bennett},
  {Gold}, {Hinshaw}, {Jarosik}, {Larson}, {Nolta}, {Page}, {Spergel},
  {Halpern}, {Hill}, {Kogut}, {Limon}, {Meyer}, {Odegard}, {Tucker}, {Weiland},
  {Wollack}, \& {Wright}}]{kom11}
{Komatsu}, E., {Smith}, K.~M., {Dunkley}, J., et al. 2011, ApJS, 192, 18

\bibitem[{{Kuhn} {et~al.}(2001){Kuhn}, {Elvis}, {Bechtold}, \&
  {Elston}}]{kuh01}
{Kuhn}, O., {Elvis}, M., {Bechtold}, J., \& {Elston}, R. 2001, ApJS, 136, 225

\bibitem[{{Kurk} {et~al.}(2007){Kurk}, {Walter}, {Fan}, {Jiang}, {Riechers},
  {Rix}, {Pentericci}, {Strauss}, {Carilli}, \& {Wagner}}]{kur07}
{Kurk}, J.~D., {Walter}, F., {Fan}, X., et al. 2007, ApJ, 669, 32

\bibitem[{{Latif} {et~al.}(2013){Latif}, {Schleicher}, {Schmidt}, \&
  {Niemeyer}}]{lat13}
{Latif}, M.~A., {Schleicher}, D.~R.~G., {Schmidt}, W., \& {Niemeyer}, J. 2013,
  MNRAS, in press (arXiv:1304.0962)

\bibitem[{{Lawrence} {et~al.}(2007){Lawrence}, {Warren}, {Almaini}, {Edge},
  {Hambly}, {Jameson}, {Lucas}, {Casali}, {Adamson}, {Dye}, {Emerson},
  {Foucaud}, {Hewett}, {Hirst}, {Hodgkin}, {Irwin}, {Lodieu}, {McMahon},
  {Simpson}, {Smail}, {Mortlock}, \& {Folger}}]{law07}
{Lawrence}, A., {Warren}, S.~J., {Almaini}, O., et al. 2007, MNRAS, 379, 1599

\bibitem[{{Lewis} {et~al.}(2010){Lewis}, {Irwin}, \& {Bunclark}}]{lew10}
{Lewis}, J.~R., {Irwin}, M., \& {Bunclark}, P. 2010, in Astronomical Society of
  the Pacific Conference Series, Vol. 434, Astronomical Data Analysis Software
  and Systems XIX, ed. Y.~{Mizumoto}, K.-I. {Morita}, \& M.~{Ohishi}, 91

\bibitem[{{Lewis} {et~al.}(2005){Lewis}, {Irwin}, {Hodgkin}, {Bunclark},
  {Evans}, \& {McMahon}}]{lew05}
{Lewis}, J.~R., {Irwin}, M.~J., {Hodgkin}, S.~T., et al. 2005, in Astronomical
Society of the Pacific Conference Series, Vol. 347, Astronomical Data Analysis
Software and Systems XIV, ed. P.~{Shopbell}, M.~{Britton}, \& R.~{Ebert}, 599

\bibitem[{{McGreer} {et~al.}(2013){McGreer}, {Jiang}, {Fan}, {Richards},
  {Strauss}, {Ross}, {White}, {Shen}, {Schneider}, {Myers}, {Niel Brandt},
  {DeGraf}, {Glikman}, {Ge}, \& {Streblyanska}}]{mcg13}
{McGreer}, I.~D., {Jiang}, L., {Fan}, X., et al. 2013, ApJ, 768, 105

\bibitem[{{Modigliani} {et~al.}(2010){Modigliani}, {Goldoni}, {Royer},
  {Haigron}, {Guglielmi}, {Fran{\c c}ois}, {Horrobin}, {Bristow}, {Vernet},
  {Moehler}, {Kerber}, {Ballester}, {Mason}, \& {Christensen}}]{mod10}
{Modigliani}, A., {Goldoni}, P., {Royer}, F., et al. 2010, in
  Society of Photo-Optical Instrumentation Engineers (SPIE) Conference Series,
  Vol. 7737, Society of Photo-Optical Instrumentation Engineers (SPIE)
  Conference Series

\bibitem[{{Morganson} {et~al.}(2012){Morganson}, {De Rosa}, {Decarli},
  {Walter}, {Chambers}, {McGreer}, {Fan}, {Burgett}, {Flewelling}, {Greiner},
  {Hodapp}, {Kaiser}, {Magnier}, {Price}, {Rix}, {Sweeney}, \&
  {Waters}}]{mor12}
{Morganson}, E., {De Rosa}, G., {Decarli}, R., et al. 2012, AJ, 143, 142

\bibitem[{{Mortlock} {et~al.}(2011){Mortlock}, {Warren}, {Venemans}, {Patel},
  {Hewett}, {McMahon}, {Simpson}, {Theuns}, {Gonz{\'a}les-Solares}, {Adamson},
  {Dye}, {Hambly}, {Hirst}, {Irwin}, {Kuiper}, {Lawrence}, \&
  {R{\"o}ttgering}}]{mor11}
{Mortlock}, D.~J., {Warren}, S.~J., {Venemans}, B.~P., et al. 2011,
  Nature, 474, 616

\bibitem[{{Richards} {et~al.}(2002{\natexlab{a}}){Richards}, {Fan}, {Newberg},
  {Strauss}, {Vanden Berk}, {Schneider}, {Yanny}, {Boucher}, {Burles},
  {Frieman}, {Gunn}, {Hall}, {Ivezi{\'c}}, {Kent}, {Loveday}, {Lupton},
  {Rockosi}, {Schlegel}, {Stoughton}, {SubbaRao}, \& {York}}]{ric02a}
{Richards}, G.~T., {Fan}, X., {Newberg}, H.~J., et al. 2002{\natexlab{a}},
  AJ, 123, 2945

\bibitem[{{Richards} {et~al.}(2002{\natexlab{b}}){Richards}, {Vanden Berk},
  {Reichard}, {Hall}, {Schneider}, {SubbaRao}, {Thakar}, \& {York}}]{ric02b}
{Richards}, G.~T., {Vanden Berk}, D.~E., {Reichard}, T.~A., et al.
  2002{\natexlab{b}}, AJ, 124, 1

\bibitem[{{Schmidt} {et~al.}(1995){Schmidt}, {Schneider}, \& {Gunn}}]{sch95}
{Schmidt}, M., {Schneider}, D.~P., \& {Gunn}, J.~E. 1995, AJ, 110, 68

\bibitem[{{Schneider} {et~al.}(2007){Schneider}, {Hall}, {Richards}, {Strauss},
  {Vanden Berk}, {Anderson}, {Brandt}, {Fan}, {Jester}, {Gray}, {Gunn},
  {SubbaRao}, {Thakar}, {Stoughton}, {Szalay}, {Yanny}, {York}, {Bahcall},
  {Barentine}, {Blanton}, {Brewington}, {Brinkmann}, {Brunner}, {Castander},
  {Csabai}, {Frieman}, {Fukugita}, {Harvanek}, {Hogg}, {Ivezi{\'c}}, {Kent},
  {Kleinman}, {Knapp}, {Kron}, {Krzesi{\'n}ski}, {Long}, {Lupton}, {Nitta},
  {Pier}, {Saxe}, {Shen}, {Snedden}, {Weinberg}, \& {Wu}}]{sch07}
{Schneider}, D.~P., {Hall}, P.~B., {Richards}, G.~T., et al. 2007, AJ,
  134, 102

\bibitem[{{Simcoe} {et~al.}(2008){Simcoe}, {Burgasser}, {Bernstein}, {Bigelow},
  {Fishner}, {Forrest}, {McMurtry}, {Pipher}, {Schechter}, \& {Smith}}]{sim08}
{Simcoe}, R.~A., {Burgasser}, A.~J., {Bernstein}, R.~A., et al. 2008, in
Society of Photo-Optical Instrumentation Engineers (SPIE) Conference Series,
Vol. 7014, Society of Photo-Optical Instrumentation Engineers (SPIE)
Conference Series

\bibitem[{{Simcoe} {et~al.}(2010){Simcoe}, {Burgasser}, {Bochanski},
  {Schechter}, {Bernstein}, {Bigelow}, {Pipher}, {Forrest}, {McMurtry},
  {Smith}, \& {Fishner}}]{sim10}
{Simcoe}, R.~A., {Burgasser}, A.~J., {Bochanski}, J.~J., et al. 2010, in
Society of Photo-Optical Instrumentation Engineers (SPIE) Conference Series,
Vol. 7735, Society of Photo-Optical Instrumentation Engineers (SPIE)
Conference Series

\bibitem[{{Simcoe} {et~al.}(2012){Simcoe}, {Sullivan}, {Cooksey}, {Kao},
  {Matejek}, \& {Burgasser}}]{sim12}
{Simcoe}, R.~A., {Sullivan}, P.~W., {Cooksey}, K.~L., et al. 2012, Nature,
492, 79

\bibitem[{{Songaila}(2004)}]{son04}
{Songaila}, A. 2004, AJ, 127, 2598

\bibitem[{{Vacca} {et~al.}(2003){Vacca}, {Cushing}, \& {Rayner}}]{vac03}
{Vacca}, W.~D., {Cushing}, M.~C., \& {Rayner}, J.~T. 2003, PASP, 115, 389

\bibitem[{{Venemans} {et~al.}(2012){Venemans}, {McMahon}, {Walter}, {Decarli},
  {Cox}, {Neri}, {Hewett}, {Mortlock}, {Simpson}, \& {Warren}}]{ven12}
{Venemans}, B.~P., {McMahon}, R.~G., {Walter}, F., et al. 2012, ApJL, 751, L25

\bibitem[{{Venemans} {et~al.}(2007){Venemans}, {McMahon}, {Warren},
  {Gonzalez-Solares}, {Hewett}, {Mortlock}, {Dye}, \& {Sharp}}]{ven07b}
{Venemans}, B.~P., {McMahon}, R.~G., {Warren}, S.~J., et al. 2007,
  MNRAS, 376, L76

\bibitem[{{Vernet} {et~al.}(2011){Vernet}, {Dekker}, {D'Odorico}, {Kaper},
  {Kjaergaard}, {Hammer}, {Randich}, {Zerbi}, {Groot}, {Hjorth}, {Guinouard},
  {Navarro}, {Adolfse}, {Albers}, {Amans}, {Andersen}, {Andersen}, {Binetruy},
  {Bristow}, {Castillo}, {Chemla}, {Christensen}, {Conconi}, {Conzelmann},
  {Dam}, {de Caprio}, {de Ugarte Postigo}, {Delabre}, {di Marcantonio},
  {Downing}, {Elswijk}, {Finger}, {Fischer}, {Flores}, {Fran{\c c}ois},
  {Goldoni}, {Guglielmi}, {Haigron}, {Hanenburg}, {Hendriks}, {Horrobin},
  {Horville}, {Jessen}, {Kerber}, {Kern}, {Kiekebusch}, {Kleszcz}, {Klougart},
  {Kragt}, {Larsen}, {Lizon}, {Lucuix}, {Mainieri}, {Manuputy}, {Martayan},
  {Mason}, {Mazzoleni}, {Michaelsen}, {Modigliani}, {Moehler}, {M{\o}ller},
  {Norup S{\o}rensen}, {N{\o}rregaard}, {P{\'e}roux}, {Patat}, {Pena}, {Pragt},
  {Reinero}, {Rigal}, {Riva}, {Roelfsema}, {Royer}, {Sacco}, {Santin},
  {Schoenmaker}, {Spano}, {Sweers}, {Ter Horst}, {Tintori}, {Tromp}, {van
  Dael}, {van der Vliet}, {Venema}, {Vidali}, {Vinther}, {Vola}, {Winters},
  {Wistisen}, {Wulterkens}, \& {Zacchei}}]{ver11}
{Vernet}, J., {Dekker}, H., {D'Odorico}, S., et al. 2011, A\&A, 536, A105

\bibitem[{{Vernet} {et~al.}(2010){Vernet}, {Kerber}, {Mainieri}, {Rauch},
  {Saitta}, {D'Odorico}, {Bohlin}, {Ivanov}, {Lidman}, {Mason}, {Smette},
  {Walsh}, {Fosbury}, {Goldoni}, {Groot}, {Hammer}, {Kaper}, {Horrobin},
  {Kjaergaard-Rasmussen}, \& {Royer}}]{ver10}
{Vernet}, J., {Kerber}, F., {Mainieri}, V., et al. 2010, Highlights of
Astronomy, 15, 535

\bibitem[{{Vestergaard} \& {Osmer}(2009)}]{ves09}
{Vestergaard}, M. \& {Osmer}, P.~S. 2009, ApJ, 699, 800

\bibitem[{{Vestergaard} \& {Wilkes}(2001)}]{ves01}
{Vestergaard}, M. \& {Wilkes}, B.~J. 2001, ApJS, 134, 1

\bibitem[{{Walter} {et~al.}(2009){Walter}, {Riechers}, {Cox}, {Neri},
  {Carilli}, {Bertoldi}, {Weiss}, \& {Maiolino}}]{wal09b}
{Walter}, F., {Riechers}, D., {Cox}, P., et al. 2009, Nature, 457, 699

\bibitem[{{Wang} {et~al.}(2013){Wang}, {Wagg}, {Carilli}, {Walter}, {Lentati},
    {Fan}, {Riechers}, {Bertoldi}, {Narayanan}, {Strauss}, {Cox}, {Omont},
    {Menten}, {Knudsen}, {Neri}, \& {Jiang}}]{wan13} 
{Wang}, R., {Wagg}, J., {Carilli}, C.~L., et al. 2013, ApJ, accepted
  (arXiv:1302.4154)

\bibitem[{{Willott} {et~al.}(2010{\natexlab{a}}){Willott}, {Albert},
  {Arzoumanian}, {Bergeron}, {Crampton}, {Delorme}, {Hutchings}, {Omont},
  {Reyl{\'e}}, \& {Schade}}]{wil10b}
{Willott}, C.~J., {Albert}, L., {Arzoumanian}, D., et al. 2010{\natexlab{a}},
AJ, 140, 546

\bibitem[{{Willott} {et~al.}(2005){Willott}, {Delfosse}, {Forveille},
  {Delorme}, \& {Gwyn}}]{wil05}
{Willott}, C.~J., {Delfosse}, X., {Forveille}, T., {Delorme}, P., \& {Gwyn},
  S.~D.~J. 2005, ApJ, 633, 630

\bibitem[{{Willott} {et~al.}(2007){Willott}, {Delorme}, {Omont}, {Bergeron},
  {Delfosse}, {Forveille}, {Albert}, {Reyl{\'e}}, {Hill}, {Gully-Santiago},
  {Vinten}, {Crampton}, {Hutchings}, {Schade}, {Simard}, {Sawicki}, {Beelen},
  \& {Cox}}]{wil07}
{Willott}, C.~J., {Delorme}, P., {Omont}, A., et al. 2007, AJ, 134, 2435

\bibitem[{{Willott} {et~al.}(2010{\natexlab{b}}){Willott}, {Delorme},
  {Reyl{\'e}}, {Albert}, {Bergeron}, {Crampton}, {Delfosse}, {Forveille},
  {Hutchings}, {McLure}, {Omont}, \& {Schade}}]{wil10a}
{Willott}, C.~J., {Delorme}, P., {Reyl{\'e}}, C., et al. 2010{\natexlab{b}},
AJ, 139, 906

\bibitem[{{Worseck} \& {Prochaska}(2011)}]{wor11}
{Worseck}, G. \& {Prochaska}, J.~X. 2011, ApJ, 728, 23

\end{thebibliography}
\end{document}